\documentclass[sigconf]{acmart}

\AtBeginDocument{%
  }

\copyrightyear{2023}
\acmYear{2023}
\setcopyright{rightsretained}
\acmConference[IMC '23]{Proceedings of the 2023 ACM Internet Measurement Conference}{October 24--26, 2023}{Montreal, QC, Canada}
\acmBooktitle{Proceedings of the 2023 ACM Internet Measurement Conference (IMC '23), October 24--26, 2023, Montreal, QC, Canada}
\acmDOI{10.1145/3618257.3624803}
\acmISBN{979-8-4007-0382-9/23/10}

\makeatletter
\gdef\@copyrightpermission{
  \begin{minipage}{0.3\columnwidth}
   \href{https://creativecommons.org/licenses/by/4.0/}{\includegraphics[width=0.90\textwidth]{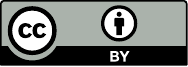}}
  \end{minipage}\hfill
  \begin{minipage}{0.7\columnwidth}
   \href{https://creativecommons.org/licenses/by/4.0/}{This work is licensed under a Creative Commons Attribution International 4.0 License.}
  \end{minipage}
  \vspace{5pt}
}
\makeatother



\usepackage{url}
\usepackage{booktabs}
\usepackage{xspace}
\usepackage{xurl}  
\usepackage{subfig}
\usepackage{multirow}
\usepackage{soul, color}
\usepackage{enumitem}
\usepackage{tikz}
\usepackage{listings}
\usepackage[flushleft]{threeparttable}
\usepackage{eurosym}
\usepackage{colortbl}
\usepackage{xcolor}
\usepackage{makecell}
\usepackage{csquotes}
\usepackage{adjustbox}
\usepackage{balance}
\usepackage{comment}
\usepackage{threeparttable}
\usepackage{hyperref}

\newcolumntype{?}[1]{!{\vrule width #1}}

\newcommand{\echo}{Alexa Echo}

\providecommand{\ione}{\emph{(i)}}
\providecommand{\itwo}{\emph{(ii)}}
\providecommand{\ithree}{\emph{(iii)}}

\newcommand{\green}[1]{\colorbox[HTML]{8adba0}{#1}}
\newcommand{\yellow}[1]{\colorbox[HTML]{fcd14f}{#1}}
\newcommand{\red}[1]{\colorbox[HTML]{ff9999}{#1}}
\newcommand{\gray}[1]{\colorbox[HTML]{C0C0C0}{#1}}

\newcommand{\greenbox}[1]{\adjustbox{margin=2px,bgcolor={HTML}{8adba0},padding=1px}{#1}}
\newcommand{\yellowbox}[1]{\adjustbox{margin=2px,bgcolor={HTML}{fcd14f},padding=1px}{#1}}
\newcommand{\redbox}[1]{\adjustbox{margin=2px,bgcolor={HTML}{ff9999},padding=1px}{#1}}
\newcommand{\graybox}[1]{\adjustbox{margin=2px,bgcolor={HTML}{c0c0c0},padding=1px}{#1}}

\begin{document}

\title{Tracking, Profiling, and Ad Targeting in the Alexa Echo Smart Speaker Ecosystem}

\author{Umar Iqbal}
\affiliation{%
  \institution{Washington University in St. Louis}
  \country{}
}

\author{Pouneh Nikkhah Bahrami}
\affiliation{%
  \institution{University of California, Davis}
  \country{}
}

\author{Rahmadi Trimananda}
\affiliation{%
  \institution{University of California, Irvine}
  \country{}
}

\author{Hao Cui}
\affiliation{%
  \institution{University of California, Irvine}
  \country{}
}

\author{Alexander Gamero-Garrido}
\affiliation{%
  \institution{University of California, Davis}
  \country{}
}

\author{Daniel J. Dubois}
\affiliation{%
  \institution{Northeastern University}
  \country{}
}

\author{David Choffnes}
\affiliation{%
  \institution{Northeastern University}
  \country{}
}

\author{Athina Markopoulou}
\affiliation{%
  \institution{University of California, Irvine}
  \country{}
}

\author{Franziska Roesner}
\affiliation{%
  \institution{University of Washington}
  \country{}
}

\author{Zubair Shafiq}
\affiliation{%
  \institution{University of California, Davis}
  \country{}
}

\renewcommand{\shortauthors}{Umar Iqbal et al.}

\renewcommand{\authors}{Umar Iqbal, Pouneh Nikkhah Bahrami, Rahmadi Trimananda, Hao Cui, Alexander Gamero-Garrido, Daniel J. Dubois, David Choffnes, Athina Mar-kopoulou, Franziska Roesner, and Zubair Shafiq}

\begin{abstract}
Smart speakers collect voice commands, which can be used to infer sensitive information about users. 
Given the potential for privacy harms, there is a need for greater transparency and control over the data collected, used, and shared by smart speaker platforms as well as third party skills supported on them. 
To bridge this gap, we build a framework to measure data collection, usage, and sharing by the smart speaker platforms. 
We apply our framework to the Amazon smart speaker ecosystem.
Our results show that Amazon and third parties, including advertising and tracking services that are unique to the smart speaker ecosystem, collect smart speaker interaction data.
We also find that Amazon processes smart speaker interaction data to infer user interests and uses those inferences to serve targeted ads to users.
Smart speaker interaction also leads to ad targeting and as much as 30$\times$ higher bids in ad auctions, from third party advertisers.
Finally, we find that Amazon's and third party skills' data practices are often not clearly disclosed in their policy documents.
\end{abstract}

\maketitle

\section{Introduction}
\label{sec:introduction}

Smart speakers are becoming increasingly prevalent because of their convenience \cite{smart_speaker_usage_global}. 
However, with that convenience come privacy risks, since the data collected by smart speakers (i.e., voice recordings, their transcripts, and interaction metadata) can reveal, or can be used to infer, sensitive information about the users. 
For example, smart speaker vendors or third parties may infer users' sensitive physical (e.g., age, health) and psychological (e.g., mood, confidence) traits from their voice recordings~\cite{Singh2019VoiceProfiling}.
Similarly, the set of questions and commands issued to a smart speaker may reveal sensitive information about users' states of mind, interests, and concerns. 
Despite the significant potential for privacy harms, users have little-to-no visibility into what information is captured by smart speakers, how it is shared with other parties, or how it is used by such parties.

Prior work provides ample evidence to support the need for greater transparency into smart speaker data collection, usage, and sharing. 
For instance, smart speaker platforms have been previously found to host malicious third party apps \cite{Cheng2020SkillPolciyViolationCCS, Young22SkillDetectiveUSENIX}, to mis-activate~\cite{dubois2020speakersPETS}, and record users' private conversations without their knowledge \cite{smart_speaker_eavesdropping, alexa_misactivation_real_world}, and to share users' conversations with strangers \cite{alexa_data_shared_with_stranger}.
There is a clear need to audit how smart speaker ecosystems handle data from their users' interactions. 

To facilitate such independent, repeatable audits, we need an approach that can work on unmodified, off-the-shelf devices, without relying on cooperation by the smart speaker manufacturer. 
Conducting such audits requires addressing two key open challenges.
First, commercially available smart speakers are closed-box devices without open interfaces that would allow independent researchers to expose what data is collected. 
Second, once data gathered from a smart speaker is sent over the Internet, there is no way to track how the data is further shared and used. 

In this paper, we address these challenges by building an auditing framework that measures the collection, usage, and sharing of smart speaker interaction data.
\emph{Our key insight is that data collection, usage, and sharing can be: (i)  directly observed by intercepting the network traffic on the router without modifying smart speakers and (ii) indirectly inferred through its usage in targeted advertisements.}
To operationalize this insight, we first expose data to smart speakers, observe the endpoints contacted during that exposure to capture the online services that collect data, and then analyze the targeted advertisements to infer the usage and sharing of the exposed data.
We apply our framework to the Alexa Echo smart speaker ecosystem, which is the largest ecosystem, with 46 million devices in the US \cite{echo_installed_customer_base} and more than 200K third party applications \cite{edu2021skillvet}.

To expose data, we simulate several treatment and control \textit{personas} with different smart speaker usage profiles. 
Each treatment persona is simulated by installing and interacting with \textit{skills} (the term for apps in the Alexa Echo ecosystem) from different categories on separate Alexa Echos, according to personas (e.g., a ``fashion'' persona is configured by installing and interacting with skills from the fashion category). 
By contrast, in the control persona, we do not install and interact with skills on Alexa Echo, and consequently do not expose any data.

To measure data collection, we build a custom Raspberry Pi (RPi) router \cite{rpi_access_point} that allows us to capture the network endpoints contacted, by unmodified off-the-shelf \echo{}s, during the skill installation and interaction process.
To infer data usage and sharing, we look for statistically significant differences in the online targeted advertising between the treatment and control personas~\cite{Papadopoulos17IMCYouAreTheProduct,Cook20HeaderBiddingPETS,Bashir16TrackingFlowsUsenix}.
We measure targeting across two modalities: ad auction bid values and ad content served to the personas. 
By comparing ad auction bid values and ad content across treatment (i.e., when data is exposed) and control (i.e., when data is not exposed) personas, we can identify when smart speaker interactions are likely the cause of ad targeting, and thus infer that the data was used and/or shared for that purpose. 

We summarize our key contributions as follows:

\begin{enumerate}
    \item We develop a novel framework, that relies on direct and indirect measurements to understand the data collection, usage, and sharing in the Alexa Echo smart speaker ecosystem, without relying on cooperation from the smart speaker manufacturer. 
    \item We find that Alexa Echo interaction data is collected by both Amazon and third parties, including advertising and tracking services. As many as 41 advertisers sync their cookies (i.e., share data) with Amazon, and further with 247 other third parties, including advertising services.
    \item We find evidence that Amazon processes Echo interactions to infer user interests, which was not clearly stated in Amazon's policies before our research and public disclosure. 
    Our measurements also indicate that inferred interests are used to serve targeted ads to users on the web. Advertisers bid as much as 30$\times$ higher on some personas. 
    \item Third party skills often do not clearly disclose their data collection practices in their privacy policies. For example, only 10 (2.2\%) skills clearly disclose the endpoints that collect data and 68.61\% (129/188) of skills do not even mention Alexa or Amazon in their privacy policies.
\end{enumerate}

\textbf{Paper organization.}The rest of this paper is organized as follows: Section \ref{sec:background} provides background and motivation behind our research. 
Section \ref{sec:methodology} presents our proposed auditing framework.
Section \ref{sec:network-traffic-analysis} presents the results of network traffic analysis to measure data collection. 
Section \ref{sec:data-usage-and-sharing-analysis} presents the ad targeting analysis to infer data usage and sharing. 
Section \ref{sec:privacy-policy-analysis} analyzes the consistency of privacy policies, and other disclosures, with observed data collection, usage, and sharing practices. 
Section \ref{sec:discussion} provides discussion and Section \ref{sec:conclusion} concludes the paper. 

\section{Background \& Motivation}
\label{sec:background}

\subsection{Alexa \& Echo}
In this paper, we study Alexa Echo smart speaker ecosystem, the most widely used ecosystem with more than 46 million devices in the US \cite{echo_installed_customer_base}.
Echos are Alexa-powered smart speakers from Amazon. 
Alexa is a voice assistant that responds to user requests conveyed through voice input. 
Although Alexa can respond to a wide variety of general-purpose requests, it is not well-suited for specialized tasks, e.g., ordering a pizza from a particular restaurant. 
Thus, to augment Alexa, Amazon allows third party services to 
build and publish applications called \textit{skills} on the Alexa marketplace. 
Alexa marketplace hosts more than 200K third party skills~\cite{edu2021skillvet}.

\subsection{Privacy issues}
The inclusion of third party skills poses a privacy risk to the users of Alexa Echos.
Accordingly, Amazon imposes a set of platform policies to mitigate potential privacy risks of third party skills.
Amazon restricts skills from collecting sensitive information, e.g., social security and bank account numbers~\cite{amazon_skills_privacy_policy, amazon_skills_policy}, and requires user permission to allow access to personal information, e.g., email, phone, location~\cite{amazon_skill_permission_model}.
To enforce the aforementioned policies, Amazon has a skill certification process that aims to filter malicious skills before they can be published on the marketplace~\cite{amazon_skill_certification_process}. 
However, prior research has shown that policy-violating skills can get certified~\cite{Cheng2020SkillPolciyViolationCCS} and thousands of skills on the Alexa marketplace violate platform policies~\cite{Young22SkillDetectiveUSENIX}.

Smart speaker platforms, such as the Alexa Echo, also store voice recordings, their transcripts, and the information about the resulting action (metadata) generated from the user's voice commands.
This data can be used to infer sensitive information about the user. 
For example, user commands in their raw form, i.e., voice recording, can be used to infer several physical (e.g., age, health) and psychological characteristics (e.g., mood, confidence) of the user~\cite{Singh2019VoiceProfiling}. 
User commands in their processed form, i.e., transcripts of voice recordings, can expose sensitive information (e.g., private conversations) about the user. 
Even the metadata that results in execution of a  user command (e.g., a user querying a third party health-related skill) can leak sensitive information about the user. 
Amazon aims to limit some of these privacy issues through its platform design choices~\cite{alexa_privacy_hub}.
For example, to avoid recording private conversations, user commands are only recorded when a user utters the \textit{wake word}, e.g., Alexa (though prior research has also shown that smart speakers often \textit{misactivate} and unintentionally record conversations~\cite{dubois2020speakersPETS}).
Additionally, Amazon currently does not directly use users' voice recordings to serve targeted advertisements \cite{nyt_alexa_ads}, despite patenting that idea~\cite{jin2018voice-Amazon-Patent}.
Further, only keywords from transcripts of user commands are shared with third-party skills, instead of the raw audio~\cite{alexa_skill_design_docs}.

The limiting of voice recordings and their full transcripts reduce the privacy risks posed to users but unfortunately do not fully eliminate them.
For example, Amazon and third parties can still use the metadata generated through the voice commands to profile users, and then serve them targeted advertisements. 
Further, the metadata could also be shared with other parties, a common practice in several IoT platforms~\cite{mazhar2020characterizing,huang2020iot,ren-imc19, saidi-imc20,mandalari-pets21,girish-imc23,hu-imc23}. 
These practices are particularly concerning in context of third parties because neither users nor Amazon have any visibility or control over the processing, sharing, and selling of user data; worse, third party skills often do not publish their privacy policies, nor adhere to them even when they do~\cite{edu2021skillvet}. 

Thus, in this paper, we focus on the collection, usage, and sharing of metadata, that is generated through installing and interacting with skills, and refer to it as \textit{smart speaker interaction data} or \textit{user data}, throughout the paper.

\subsection{Research questions}
To the best of our knowledge, prior work lacks an in-depth analysis of the collection, usage, and sharing of user data in smart speaker ecosystems. 
To fill this gap, we systematically analyze the data collection, usage, and sharing practices in Alexa Echo smart speaker platform, including third party skills. 
We conduct controlled experiments where we intentionally expose data by installing and interacting with skills and observe platform's behavior from three perspectives: 
\ione{} \textit{network traffic} exchanged by smart speakers, \itwo{} \textit{advertisements} served to smart speaker users, and \ithree{} \textit{privacy policies} published by third party skills.
Our goal is to combine these perspectives to answer the following research questions.

\textit{RQ1: Which organizations collect and propagate user data?}   
We use the remote endpoints of network traffic to measure data collection and sharing by Amazon and third party skills.
We can intercept and observe communication between Amazon and some (but not all) third parties; however, the Amazon ecosystem does not provide interfaces to facilitate comprehensive analysis of data collection, usage, and sharing. This motivates the need for inference below.

\textit{RQ2: Is smart speaker interaction data used by either Amazon or third party skills beyond purely functional purposes, such as for targeted advertising?} 
We measure advertisements to infer data usage and sharing by Amazon and third party skills.
To this end, we focus on detecting behaviorally targeted web ads.
We study targeting in web ads because web publishers almost universally employ well-established programmatic advertising protocols~\cite{HB_protocol, RTB_protocol}.\footnote{We do not directly study ads served on Echos because the Alexa advertising ecosystem is relatively nascent with several restrictions.
For example, Amazon only allows audio ads on streaming skills \cite{amazon_advertising_restrictions} and typically requires rather high minimum ad spend commitment from advertisers \cite{amazon_audioads_budget}.}

\textit{RQ3: Are data usage and sharing practices compliant with privacy policies and other disclosures?}
We extract key elements from privacy policies of Amazon and third party skills. 
For Amazon, we also review Alexa specific policies in Alexa Privacy Hub \cite{alexa_privacy_hub} and Alexa Device FAQs \cite{alexa_device_faq}.
We also compare policy disclosures with our measurements and inferences to assess the compliance of data collection, usage, and sharing.

\section{Auditing Framework}
\label{sec:methodology}

\begin{figure}[!t]
    \centering
    \includegraphics[width=\columnwidth]{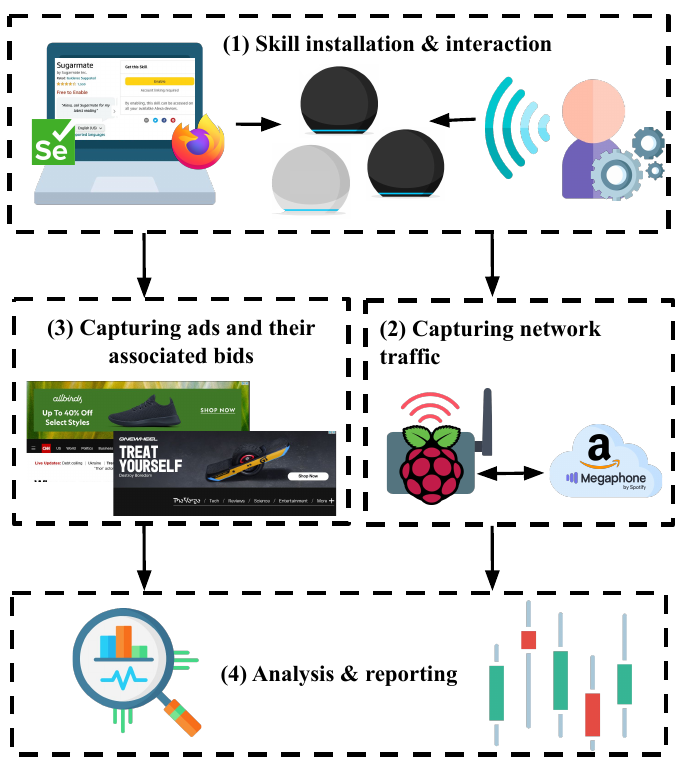}
    \caption{Approach overview: (1) We install and interact with skills from 9 different categories on 9 different smart speakers to train 9 smart speaker interest personas. (2) While installing and interacting, we also capture and store network traffic to/from \echo{}. In addition to interest (treatment) personas, we also train a vanilla (control) persona where we do not install and interact with skills. (3) We then visit popular websites while logged into each persona's Amazon account to capture and store ads and their associated bids targeted to the personas. (4) We then analyze this data to measure data collection, usage, sharing, and its compliance with Amazon's and skills policies.}
    \label{fig:va-approach}
\end{figure}

In this section, we describe our methodology for measuring data collection, usage, and sharing of user interaction data by Amazon and third party skills.
Figure \ref{fig:va-approach} presents the overview of our approach.
At a high level, (1) we intentionally expose data by installing and interacting with skills on Alexa Echos; (2) capture the endpoints that collect this data by intercepting network traffic; and (3) then analyze ad targeting on popular websites to infer the usage of exposed data. 
These steps are described next, in Sections \ref{sec:leaking-data}, \ref{sec:capturing-network-traffic}, and \ref{subsection:capturing-advertisements}, respectively.
Our experiments were conducted from University of Washington and University of California, Davis in the US. 
All of the IP addresses for respective experiments geolocated to the same approximate location.

\subsection{Simulating personas to expose data}
\label{sec:leaking-data}

\subsubsection{Simulating interest personas.}\label{sec:simulating-skill-personas}
We simulate nine interest (treatment) personas by installing and interacting with skills from nine different Alexa marketplace categories: Connected Car, Dating, Fashion \& Style, Pets \& Animals, Religion \& Spirituality, Smart Home, Wine \& Beverages, Health \& Fitness, and Navigation \& Trip Planners.
We simulate several personas because the nature of tracking, profiling, and ad targeting might differ across different skill categories.
We refer to each interest persona according to its corresponding skill category name, and collectively as \textit{interest personas}.

\textit{Skill installation.} As a first step, we create dedicated Amazon accounts for each persona and use them to configure Alexa Echos (4th generation Alexa Echo smart speakers).
To avoid contamination across personas, we configure each device through a fresh browser profile and assign it a unique IP address (all IPs geolocate to the same location).  
We then use a Selenium-based~\cite{selenium} web crawler to programmatically visit the Alexa skill marketplace, and iteratively install and enable the top-50 skills (based on the number of reviews) for each category.
We use the dataset released in~\cite{edu2021skillvet} to extract top skills from each category.
If prompted, we enable all of the requested permissions by a skill. 
Note that we do not link accounts for skills that require it, since this is infeasible at scale in our testbed.
Linking typically requires creating an account for the online service corresponding to the tested skill, and linking it to a physical IoT device, e.g., \textit{iRobot} skill requires to link a robot vacuum cleaner with the skill \cite{irobot_skill}.

\textit{Skill interaction.} 
After installing each skill, we interact with it by programmatically uttering sample invocations listed by each skill.\footnote{We use gTTS, a python package for Google Translate text-to-speech API~\cite{gtts_python} to convert the textual commands to utterances.}
We also parse skill descriptions to extract additional invocation utterances provided by the skill developer. 
We interact with the \echo{} by iteratively uttering each skill's invocations. 
In case Alexa expects a follow up response\footnote{If Alexa expects a follow up response and the response is not provided, Alexa asks for the response a few times. We match the last two responses to determine if a follow up response is expected.} or has a response of more than 30 seconds, e.g., playing a song, we terminate the interaction by uttering \textit{Alexa, Stop}.

\subsubsection{Simulating control persona.}
In addition to the nine interest (treatment) personas, we also simulate a control persona, referred to as \textit{vanilla} persona.
Similar to interest personas, the vanilla persona is linked to an Amazon account, an \echo{}, and a unique IP address. 
However, we do not install or interact with skills on the vanilla persona.
Vanilla persona serves as a baseline and allows to associate the deviation in the interest personas to the treatment, i.e., installation and interaction with skills, applied to them.

\subsection{Capturing network traffic to measure data collection}
\label{sec:capturing-network-traffic}
We capture outgoing and incoming network traffic, to and from, \echo{}s to measure data collection by Amazon and skills during skill installation and interaction.
Alexa Echos do not support on-device network analysis, so we intercept network traffic on the router to capture the endpoints that collect data. 
To this end, we set up a custom Raspberry Pi (RPi) based router~\cite{rpi_access_point} to intercept incoming and outgoing network traffic. 
We uninstall each skill before installing the next one, to ensure that we associate the correct network traffic to each skill.
Note that the captured network traffic is end-to-end encrypted, so we can inspect the \textit{endpoints} that collect data, but not the \textit{plaintext payloads}.

\textit{Inferring destination.} 
The captured network traffic contains the IP addresses of contacted endpoints.
We resolve these IP addresses to domain names by using the information from Domain Name System (DNS) packets in network traffic.
We further map domain names to their parent organization by leveraging information from DuckDuckGo \cite{DDGTrackerRadar}, Crunchbase \cite{Crunchbase}, and WHOIS.

\subsection{Capturing advertisements to infer data usage and sharing}
\label{subsection:capturing-advertisements}
We rely on ad content and advertisers' bidding behavior to infer data usage and sharing. 
Ad content can reveal the ad topic and consequently the user interests that advertisers might have inferred from the exposed \echo{} interaction data. 
However, ad content may lack objective or discernible association with the exposed data.  
For example, active advertising campaigns at the time of our experimentation may
 lack apparent association with the exposed data or advertising models may interpret user interests differently. 
In part to offset subjectivity, we use advertisers' bidding behavior to infer the usage and sharing of smart speaker interaction data. 
Prior research~\cite{Olejnik14SellingPrivacyNDSS,Papadopoulos17IMCYouAreTheProduct,Cook20HeaderBiddingPETS} has shown that the advertisers' bidding behavior is influenced by their pre-existing knowledge of the users, which typically results in higher bid values than cases where such knowledge is absent. 
Thus, if we encounter high bid values from advertisers, a likely cause is the usage and sharing of \echo{} interaction data.

\textit{Web advertisements.} The header bidding protocol \cite{HB_protocol} exposes bid values to client-side browser, so we collect ad bids and ad images on sites where header bidding is supported, both after skill installation and skill interaction.
To this end, we first identify top websites that support \texttt{prebid.js} \cite{prebid}, the most widely used implementation of header bidding protocol \cite{HB_usage}, and then visit those websites to capture bids and ad images. 
We extend OpenWPM \cite{Englehardt16MillionSiteMeasurementCCS} to identify and capture data on \texttt{prebid.js} supported websites.
To identify \texttt{prebid.js} supported websites, we crawl Tranco top websites list (from 09/07/2021) \cite{le2019tranco} and probe for \texttt{prebid.js} version. 
We treat a website as prebid supported if we receive a non-null \texttt{prebid.js} version. 
We stop the crawl as soon as we identify 200 prebid supported websites. 
We then crawl the \texttt{prebid.js} supported websites and intercept bidding requests (or request bids if no bid requests are made).

To more accurately simulate user behavior, we enable OpenWPM's bot mitigation and wait for 10--30 seconds between webpages.
We also visit the webpages in browser's native mode, with window size of $1366 \times 678$, instead of using the headless mode. 
Note that we crawl the \texttt{prebid.js} supported websites using the same browser profiles that are logged into the Amazon account and Alexa web companion app, and same IP addresses used to configure interest and vanilla personas (Section \ref{sec:leaking-data}).
The browser profiles and IP addresses connect personas with browsers and allow us to collect the advertisements targeted to the personas.
To make sure that we visit websites with unchanged browser profiles and to not let webpages influence profiles, we ignore the updates made by the webpages to the browser profiles.

\textit{Interpreting bids.}
In addition to user interests, advertisers consider several factors, e.g., day of the week, website popularity, to determine the bid values \cite{Olejnik14SellingPrivacyNDSS,Papadopoulos17IMCYouAreTheProduct}.
To reduce variability due to such confounding factors, we strive to keep conditions consistent across personas.
Specifically, we use identical hardware/software, collect bids at the same time (simultaneously), from the same location, and on the same websites, for all personas.
In addition, we consider only bids from ad slots that are successfully loaded across all personas, because bid values vary by ad slots \cite{Papadopoulos17IMCYouAreTheProduct} and advertisers may not bid on all ad slots across all personas.
We compare relative bid values across control and interest (treatment) personas because the absolute values can change over time, e.g., travel advertisements may get higher bids around holidays. 
Since it is non-trivial to reverse engineer and control for all the factors incorporated by advertisers, we crawl and extract bids from the  \texttt{prebid.js} supported websites several times, i.e., 6 times before interacting with skills and 25 times after interacting with skills\footnote{We terminated the experiment after 6 iterations after skill installation because we did not notice any personalization. We continued to crawl 25 times after skill interaction because we noticed personalization (Section \ref{sec:data-usage-and-sharing-analysis}) and needed more samples due to the variability in bid values.}, to further account for the variability in bid values.

\textit{Capturing requests/responses.}
In addition to collecting ad bids and images, we also record the network requests and responses while crawling popular websites. 
Network traffic allows us to measure data sharing (e.g., cookie syncing \cite{google_rtb_docs}) between Amazon and its advertising partners. 
Note that the network traffic captured while crawling (referred to as \textit{web traffic}) is separate from the network traffic captured from \echo{}s (Section \ref{sec:capturing-network-traffic}).

\section{Network Traffic Analysis}
\label{sec:network-traffic-analysis}

In this section, we analyze network traffic from the Alexa Echos. 
We identify online services that directly collect user data and also investigate the functionality offered by these services.

\begin{table}[!t]
    \centering
    \small
    \begin{tabular}{>{\bfseries}l?{0.3mm}l?{0.3mm}c}
    \toprule
    \textbf{Org.} & \textbf{Domains} & \textbf{Skills} \\
    \midrule
    \multirow{10}{*}{\textbf{\rotatebox[origin=c]{90}{Amazon}}}
    & *(11).amazon.com                                              & 895                                      \\
    & prod.amcs-tachyon.com                                           & 305                                      \\
    & api.amazonalexa.com                                             & 173                                      \\
    & *(7).cloudfront.net                                           & 146                                      \\
    & \cellcolor[HTML]{C0C0C0}device-metrics-us-2.amazon.com          & \cellcolor[HTML]{C0C0C0}123              \\
    & *(4).amazonaws.com                                            & 52                                       \\
    & acsechocaptiveportal.com                                        & 27                                       \\
    & fireoscaptiveportal.com                 & 20               \\
    & ingestion.us-east-1.prod.arteries.alexa.a2z.com                 & 7                                        \\
    & ffs-provisioner-config.amazon-dss.com                           & 2                                        \\
    
    \midrule
    
   \multirow{2}{*}{\textbf{\rotatebox[origin=c]{90}{skill}}}
   \multirow{3}{*}{\textbf{\rotatebox[origin=c]{90}{vendor}}} & *(2).youversionapi.com                                        & 2                                          \\ [1ex]
    & static.garmincdn.com                                            & 1                                          \\  [1ex]
    
    \midrule
    
    \multirow{12}{*}{\textbf{\rotatebox[origin=c]{90}{Third party}}}
    & dillilabs.com                                                    & 9                \\ 
    & \cellcolor[HTML]{C0C0C0}*(2).megaphone.fm                      & \cellcolor[HTML]{C0C0C0}9                \\ 
    & cdn2.voiceapps.com                                               & 7                                        \\
    & \cellcolor[HTML]{C0C0C0}*(2).podtrac.com                      & \cellcolor[HTML]{C0C0C0}7                \\ 
    & *(2).pod.npr.org                                               & 4                                        \\ 
    & \cellcolor[HTML]{C0C0C0}chtbl.com                                & \cellcolor[HTML]{C0C0C0}3                \\ 
    & 1432239411.rsc.cdn77.org                                         & 3                                        \\ 
    & \cellcolor[HTML]{C0C0C0}*(2).libsyn.com                        & \cellcolor[HTML]{C0C0C0}3                \\ 
    & \cellcolor[HTML]{C0C0C0}*(3).streamtheworld.com               & \cellcolor[HTML]{C0C0C0}3                \\ 
    & discovery.meethue.com                                            & 2                                        \\ 
    & \cellcolor[HTML]{C0C0C0}turnernetworksales.mc.tritondigital.com  & \cellcolor[HTML]{C0C0C0}1                \\ 
    & \cellcolor[HTML]{C0C0C0}traffic.omny.fm                          & \cellcolor[HTML]{C0C0C0}1                \\
    
    \bottomrule
    \end{tabular}
\caption{Amazon, skill vendors, and third party domains contacted by skills. ``Org.'' column refers to organization. ``Skills'' column represents the count of skills. Advertising and tracking domains are shaded with grey. Subdomain counts are represented with *(\#), e.g., *(11).amazon.com represents requests to 11 subdomains of \href{https://amazon.com}{amazon.com}.}

\label{table:domains_list}
\end{table}

\begin{figure*}[h]
    \centering
    \includegraphics[width=0.8\textwidth, trim=0cm 0cm 0cm 0.1cm, clip]{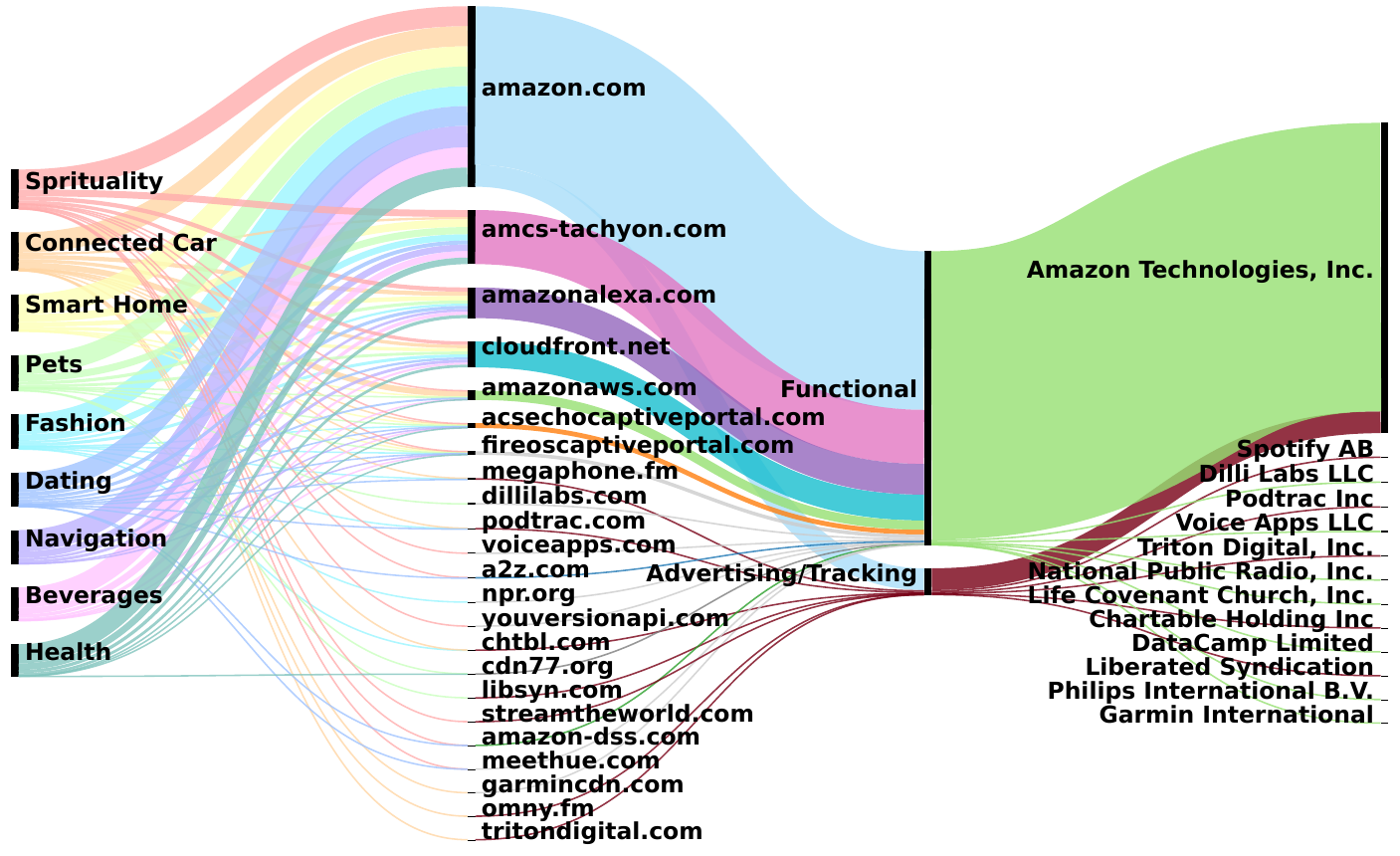}
    \caption{Network traffic distribution by persona, domain name, purpose, and organization.}
    \label{fig:sankey-bigger-version}
\end{figure*}

\subsection{Amazon as skill mediator} 
\label{sec:amazon-vantage-point}
We first analyze network traffic to identify the services that collect user data. 
Table \ref{table:domains_list} presents the list of domains contacted by skills.
We note that unlike more established platforms, e.g., mobile, the majority of the traffic in Alexa smart speaker ecosystem goes to the device manufacturer.
Specifically,  446 (99.11\%), 2 (0.45\%), and 31 (6.89\%) of the skills contact domains that belong to Amazon, skill vendors, and third parties, respectively. 
Four (0.89\%) skills failed to load.
All active skills contact Amazon because Amazon mediates communication between skills and users, i.e., Amazon first interprets the voice input and then shares it with the skill \cite{alexa_skill_design_docs}.
Another possible explanation for this is the use of Amazon's infrastructure to host skills~\cite{alexa_hosted_skills}.
\textit{Garmin} \cite{GarminSkill} and \textit{YouVersion Bible} \cite{YouVersionBibleSkill} are the only skills that send traffic to their own domains.

Figure \ref{fig:sankey-bigger-version} shows the number of network flows from skills to domains, their functionality, and their parent organizations.
Similar to results from Table \ref{table:domains_list}, most network flows involve Amazon.
We note that the skills in most categories, except for Smart Home, Wine \& Beverages, Navigation \& Trip Planners, contact third party services.

\begin{table}[h]
    \centering
    \small
    \begin{tabular}{>{\bfseries}l?{0.3mm}cc?{0.3mm}cc}
    \toprule    
    Organization       &  \textbf{Functional} & \textbf{Advertising \& Tracking}   &  \textbf{Total}\\
    \midrule
    Amazon               & 90.04\%  & 6.80\%   & 96.84\%\\
    Skill vendor          &  0.17\%  &  0\%    & 0.17\%\\
    Third party          &  1.49\%  & 1.50\%   & 2.99\%\\
    \midrule
     Total              &  91.70\%  & 8.3\%  & 100\%\\
    \bottomrule
    \end{tabular}
    \caption{Distribution of advertising / tracking and functional network traffic by organization.}
    \label{table:domain_distribution}
\end{table}

\subsection{Data shared with advertisers \& trackers}
\label{sub-sec:data-leaked-to-advertisers-and-trackers}
We next analyze the functionality offered by services that collect user data. 
Several domains contacted by skills offer audio advertising and tracking services (rows highlighted in gray in Table \ref{table:domains_list}).
We rely on filter lists \cite{PiHole} and manual analysis to detect advertising and tracking services.
Table \ref{table:domain_distribution} provides the distribution of functional and advertising domains contacted by skills.
We note that 9.4\% of all network traffic, including 1.5\% third party network traffic, supports advertising and tracking functionality.
We note that \url{device-metrics-us-2.amazon.com}, used by Amazon to collect device metrics \cite{BARCELOARMADA2022108782}, is the most prominent tracking domain.
The most contacted third party advertising and tracking services include Megaphone (\url{megaphone.fm}) and Podtrac (\url{podtrac.com}), both of which specialize in audio advertising and tracking services.
We note that prominent skills, such as \textit{Genesis} \cite{genesisskill} and \textit{Men's Finest Daily Fashion Tip} \cite{mens-finest-daily-fashion-tip-skill} with 398 and 13 reviews, contact third party advertising and tracking services. 
Despite Amazon's Alexa advertising policy restricting non-streaming skills from playing ads \cite{amazon_advertising_restrictions, amazon_advertising_restrictions_2}, we find that six non-streaming skills contact third party advertising and tracking services.
Surprisingly, we note that these skills do not play any advertisements, despite including advertising services. 
It is unclear as to why non-streaming skills include advertising and tracking services and whether such skills should be flagged during skill certification \cite{AlexaCertification}.

Tables \ref{tab:category_thirdparty} and \ref{tab:skill_rate_3pa_domains} further provide the distribution of advertising and tracking domains by personas and skills. 
From Table \ref{tab:category_thirdparty}, we note that skills in five personas contact third party advertising and tracking services, with skills in the Fashion \& Style persona contacting the most advertising and tracking services. 
Table \ref{tab:skill_rate_3pa_domains} shows that skills contact several advertising and tracking services. 
For example, the \textit{Garmin}~\cite{GarminSkill} skill contacts as many as four advertising and tracking services.

\begin{table}[t]
\centering
\small
    \begin{tabular}{>{\bfseries}l?{0.3mm}c?{0.3mm}c}
    \toprule
    Persona & \textbf{Advertising \& Tracking} & \textbf{Functional} \\

    \midrule
        Fashion \& Style           & 9 & 4  \\
        Connected Car              & 7 & 0  \\
        Pets \& Animals            & 3 & 11 \\
        Religion \& Spirituality    & 3 & 10  \\
        Dating                     & 5 & 1  \\
        Health \& Fitness           & 0 & 1  \\
        \bottomrule
    \end{tabular}
\caption{Count of advertising/tracking and functional third party domains contacted by personas.}
\label{tab:category_thirdparty}
\end{table}

\begin{table}[t]
\centering
\small
\begin{tabular}{l?{0.3mm}l}
\toprule
\textbf{Skill name}                & \textbf{Advertising \& Tracking}        \\ 
\midrule
\multirow{4}{*}{Garmin \cite{GarminSkill}}            & chtbl.com                               \\
                                   & traffic.omny.fm                         \\
                                   & dts.podtrac.com                         \\
                                   & turnernetworksales.mc.tritondigital.com \\ \midrule
\multirow{3}{*}{Makeup of the Day \cite{makeup-of-the-day-skill}} & *(2).megaphone.fm                       \\
                                   & play.podtrac.com                        \\
                                   & chtbl.com                               \\ \midrule
Men's Finest Daily             & play.podtrac.com        \\
Fashion Tip \cite{mens-finest-daily-fashion-tip-skill}                                   & *(2).megaphone.fm                       \\ 
\midrule
Dating and Relationship & play.podtrac.com       \\
Tips and advices \cite{Dating-and-Relationship-Tips-and-advices-skills}                                 & *(2).megaphone.fm                       \\ \midrule
Charles Stanley Radio \cite{CharlesStanleyRadioSkill}                                       & *(2).streamtheworld.com \\ 
\bottomrule
\end{tabular}
\caption{Top-5 skills that contact third party advertising and tracking services. Subdomain counts are represented with *(\#), e.g., *(2).megaphone.fm represents two subdomains of \href{https://megaphone.fm}{megaphone.fm}.}
\label{tab:skill_rate_3pa_domains}
\end{table}

\section{Ad Targeting Analysis}
\label{sec:data-usage-and-sharing-analysis}
In this section, we analyze whether collected data is used to profile users, as well as infer if that profiling is used in ad targeting.

\subsection{Interactions are used to infer interests}
\label{section:data-profiling}
\label{sub-sec:interest-inference-amazon}
Since Amazon allows users to access data collected about them, we request data for interest and vanilla personas \cite{amazon_request_pi}. 
The data contains detailed information about device diagnostics, search history, retail interactions, Alexa, advertising, and other Amazon services.  
We are mostly interested in advertising interests inferred by Amazon based on skill installation and interactions.
We request data thrice, once after skill installation and twice after skill interaction, to see the evolution in inferred interests over time.
Since advertising interests are inferred instantly and made available to users for download within 2 days \cite{zhang2021harpo}, we request user interests after 3 days of skill installation and 8 and 31 days of skill interaction. 
Amazon on average took around 12 days to return the inferred interests after our request.

Table \ref{table:amazon-interests} presents the advertising interests inferred by Amazon for Health \& Fitness, Fashion \& Style, and Smart Home personas. 
For remaining personas, Amazon did not return any interests. 
We note that both skill installation and interaction lead to interests inference by Amazon. 
With only skill installation, Amazon infers that Health \& Fitness persona is interested in \textit{Electronics} and \textit{DIY \& Tools}. 
Skill interaction further allows Amazon to infer interests for Fashion \& Style and Smart Home personas and also refine interests for Health \& Fitness persona.
Some of the interests inferred by Amazon seem clearly relevant to the personas.
For example, \textit{Fashion} and \textit{Beauty \& Personal Care} interests seem relevant to the Fashion \& Style persona and \textit{Home \& Kitchen} interests seem relevant to the Smart Home persona. 
Note that for our second data request after interaction, Amazon did not return advertising interest files for Health \& Fitness, Wine \& Beverages, Religion \& Spirituality, Dating, and vanilla personas---nor did they provide these files upon a third request.

Our results suggest that Amazon at the very least uses the metadata of interactions with Alexa Echo smart speakers to infer user interests for ad targeting.
This is concerning because before our research and public disclosure, these practices were not clearly stated in Amazon's policies~\cite{alexa_privacy_hub_archived, alexa_device_faq_archived} (see Section \ref{sec:amazon-disclosure} for details).

\begin{table}[t]
    \centering
    \small
    \begin{tabular}{p{1.6cm}?{0.3mm}p{2.0cm}?{0.3mm}p{3.8cm}}
    \toprule
    \textbf{Configuration} & \textbf{Persona}            & \textbf{Amazon inferred interests} \\
    \midrule
    \multirow{2}{*}{\textbf{Installation}}& \multirow{2}{*}{\textbf{Health \& Fitness}}           &  Electronics \\
    & & Home \& Garden: DIY \& Tools\\ 
    \midrule
    \multirow{7}{*}{\textbf{Interaction}} & \multirow{1}{*}{\textbf{Health \& Fitness}}           &  Home \& Garden: DIY \& Tools \\
    \cmidrule{2-3}
    \multirow{7}{*}{\textbf{(1)}}& \multirow{3}{*}{\textbf{Fashion \& Style}}    &  Beauty \& Personal Care \\
    &&Fashion \\
    &&Video Entertainment \\
    \cmidrule{2-3}
    &  \multirow{3}{*}{\textbf{Smart Home}}                  & Electronics \\
    &&Home \& Garden: DIY \& Tools \\
    &&Home \& Garden: Home \& Kitchen  \\
    
    \midrule

    \multirow{6}{*}{\textbf{Interaction}}& \multirow{2}{*}{\textbf{Fashion \& Style}}   & Fashion \\
    \multirow{6}{*}{\textbf{(2)}}&&Video Entertainment \\
    \cmidrule{2-3}
    & \multirow{3}{*}{\textbf{Smart Home}}                  & Pet Supplies \\
    &&Home \& Garden: DIY \& Tools \\
    &&Home \& Garden: Home \& Kitchen  \\

    \bottomrule
    \end{tabular}
    \caption{Advertising interests inferred by Amazon for interest personas. Installation represents advertising interest inferred after skill installation. Interaction represent advertising interests inferred after skill interaction. Data is downloaded twice after interaction, represented by (1) and (2).}
    \label{table:amazon-interests}
\end{table}

\subsection{Interactions lead to higher ad bids}
\label{sub-sec:installation-interaction-bidding}

\subsubsection{Bid values after skill installation.}
\label{sub-sec:no-interaction-bidding}
\begin{figure}[t]
    \centering
    \footnotesize
    \includegraphics[width=0.48\textwidth]{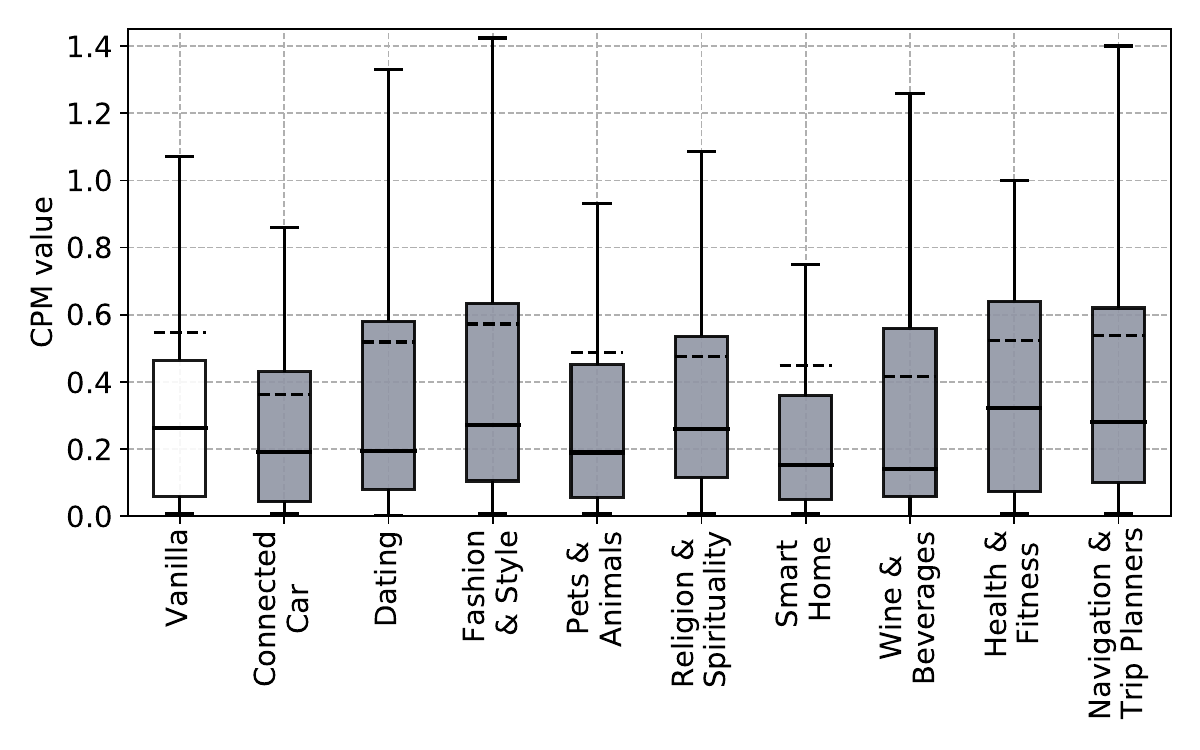}
    \caption{CPM values across vanilla (control) and interest (treatment) personas on common ad slots with skill installation. Solid and dotted lines in bars represent median and mean, respectively. Vanilla persona does not involve skill installation.}
    \label{figure:All-Websites-Common-AdSlots-Filtered-No-Interaction}
\end{figure}

To evaluate whether only skill installation leads to personalized ad targeting, we first analyze advertisers bidding behavior for vanilla and interest personas without any user interaction. 
Figure \ref{figure:All-Websites-Common-AdSlots-Filtered-No-Interaction} presents bid (CPM)\footnote{CPM (cost per mille) is the amount an advertiser pays a website per thousand visitors who see its advertisements. Bids are expressed in CPM.} values (y-axis) across vanilla and interest personas (x-axis) on common ad slots without user interaction.
The figure shows that without user interaction, there is no discernible difference between vanilla and interest personas. 
Only Fashion \& Style personas has slightly higher, statistically insignificant, mean bid value than vanilla persona. 
Overall, skill installation (without user interaction) does not lead to observable targeting.

\begin{figure}[t]
    \centering
    \footnotesize
    \includegraphics[width=0.48\textwidth]{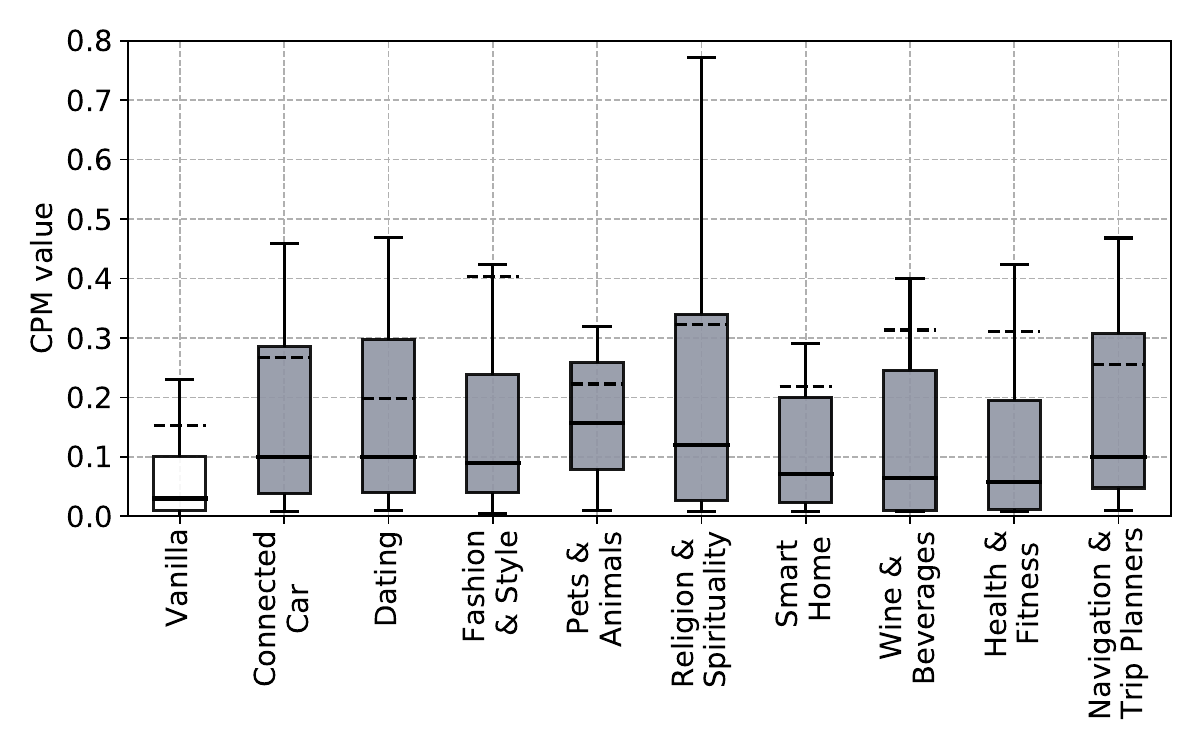}
    \caption{CPM values across vanilla (control) and interest (treatment) personas on common ad slots with user interaction. Solid and dotted lines in bars represent median and mean, respectively. Vanilla persona does not involve user interaction.}
    \label{figure:All-Websites-Common-AdSlots-Filtered-Interaction}
\end{figure}

\subsubsection{Bid values after skill installation and interaction.}
\label{sub-sec:interaction-bidding}
Next, we analyze advertisers bidding behavior for vanilla and interest personas with user interaction to evaluate if interaction with skills leads to personalized ad targeting. 
Figure \ref{figure:All-Websites-Common-AdSlots-Filtered-Interaction} presents bid (CPM) values across vanilla and interest personas on common ad slots with user interaction.
In contrast to bid values without user interaction (Figure \ref{figure:All-Websites-Common-AdSlots-Filtered-No-Interaction}), with user interaction (Figure \ref{figure:All-Websites-Common-AdSlots-Filtered-Interaction}) the bid values are significantly higher for interest personas as compared to vanilla persona. 
Table \ref{table:bids-median-mean} shows the median and mean bid values for interest and vanilla personas with user interaction. 
The table indicates that median bids for all interest personas, except for Health \& Fitness, are 2$\times$ higher than vanilla persona.
Similarly, mean bids for four interest personas, i.e., Fashion \& Style, Religion \& Spirituality, Wine \& Beverages, and Health \& Fitness, are 2$\times$ higher than vanilla persona.
We note that the bid values for Health \& Fitness and Fashion \& Style go as much as 30$\times$ and 27$\times$ higher than the mean of vanilla persona. 
It can also be seen from Table~\ref{table:bids-median-mean} that the mean bid values are higher than the median bid values, suggesting that some advertisers bid much higher than others. 
One possible explanation for this behavior could be that some advertisers have more information about the persona's interests than the others, which leads them to place much higher bids than others. 
We discuss high absolute bid values with just skill installation in Appendix \ref{appendix:high-bid-values}.

\begin{table}[t]
    \centering
    \small
    \begin{tabular}{>{\bfseries}l?{0.3mm}cc}
    \toprule
    \textbf{Persona}                     & \textbf{Median} & \textbf{Mean} \\
    \midrule
    Connected Car               & 0.099 & 0.267 \\
    Dating                      & 0.099 & 0.198 \\
    Fashion \& Style            & 0.090 & 0.403 \\
    Pets \& Animals             & 0.156 & 0.223 \\
    Religion \& Spirituality    & 0.120 & 0.323 \\
    Smart Home                  & 0.071 & 0.218 \\
    Wine \& Beverages           & 0.065 & 0.313 \\
    Health \& Fitness           & 0.057 & 0.310 \\
    Navigation \& Trip Planners & 0.099 & 0.255 \\
    \midrule
    Vanilla                     & 0.030 & 0.153 \\
    \bottomrule
    \end{tabular}
    \caption{Median and mean bid values (CPM) for interest (treatment) and vanilla (control) personas with user interaction. Vanilla persona does not involve user interaction.}
    \label{table:bids-median-mean}
\end{table}

\subsubsection{After user interaction, interest personas receive significantly higher bids.}
We perform Mann-Whitney U test to analyze whether interest personas after user interaction receive significantly higher bids than vanilla persona. 
Since we perform  multiple comparisons, we adjust our statistical significance tests with the Holm-Bonferroni correction method.
Our null hypothesis is that the bid distributions for interest personas are similar to vanilla persona.
Our alternative hypothesis is that the bid distributions for interest personas are higher than the vanilla persona.
We reject the null hypothesis when the $p$-value is less than 0.05. 
In addition to $p$-value, we also report the effect size (rank-biserial coefficient). 
Effect size ranges from -1 to 1, where -1, 0, and 1 indicate stochastic subservience, equality, and dominance of interest persona over vanilla persona. 
Effect size between 0.11--0.28, 0.28--0.43, and $\geq$ 0.43 are considered small, medium, and large, respectively.

Table \ref{table:statistical-significance} presents the results of statistical significance tests. 
We note that six interest personas have significantly higher bids than vanilla persona with medium effect size.
For the remaining three interest personas, i.e., Smart Home, Wine \& Beverages, and Health \& Fitness, the differences are not statistically significant.

\subsubsection{After user interaction, interest personas are targeted with personalized ads.}
\label{sub-sec:targeted-ads}
Next, we analyze the ads delivered through prebid.js to personas after user interaction.
In total, we receive 20,210 ads across 25 iterations.
Since ads may lack any objective or even discernible association with the shared interests, as discussed in Section \ref{subsection:capturing-advertisements}, we resort to manual analysis of ads. 
However, manual ad analysis is a tedious task and it is not feasible to analyze thousands of ads. 
To this end, we only manually analyze ads from Amazon and ads from installed skill vendors in their respective personas (e.g., an ad from Ford in Connected Car persona because it contains the \textit{FordPass} skill) because we expect these ads to be the most personalized.
We consider an ad to be personalized if it  is only present in one persona and references a product in the same industry as the installed skills (e.g., an ad for a vehicle is shown to the Connected Car persona).
While any manual labeling process is subject to human error and subjectivity, we argue that our definition is sufficiently concrete to mitigate these concerns. 
In total, we filter 79 ads from installed skills' vendors in their respective personas and 255 ads from Amazon ads for manual analysis. 
Out of the 79 ads from installed skills vendors, 60, 12, 1, and 1 are from Microsoft, SimpliSafe, Samsung, and LG in Smart Home persona, respectively. 
Out of the remaining 5, 3 are from Ford and 2 are from Jeep in Connected Car persona.
However, none of the ads from installed skills vendors are exclusive to the personas where their skills are installed, which indicates that these ads do not reveal obvious personalization.

\begin{table}[t]
    \centering
    \small
    \begin{tabular}{>{\bfseries}l?{0.3mm}ccc}
    \toprule
    \multirow{2}{*}{\textbf{Persona}}            & \multirow{2}{*}{\textbf{$p$-value}}  & \textbf{Adjusted}  & \multirow{2}{*}{\textbf{Effect size}} \\
    &  & \textbf{$p$-value} &\\

    \midrule
    \rowcolor[HTML]{C0C0C0}Connected Car               & 0.003   & 0.026  &   0.354 \\
    \rowcolor[HTML]{C0C0C0}Dating                      & 0.006   & 0.030  &  0.363 \\
    \rowcolor[HTML]{C0C0C0}Fashion \& Style            & 0.010   & 0.040  &   0.319 \\
    \rowcolor[HTML]{C0C0C0}Pets \& Animals             & 0.005   & 0.030  &   0.428 \\
    \rowcolor[HTML]{C0C0C0}Religion \& Spirituality    & 0.004   & 0.030  &   0.356 \\
    Smart Home                  & 0.075   & 0.225  &   0.210 \\
    Wine \& Beverages           & 0.083   & 0.225  &   0.192 \\
    Health \& Fitness           & 0.149   & 0.225  &   0.139 \\
    \rowcolor[HTML]{C0C0C0}Navigation \& Trip Planners & 0.002   & 0.015  &   0.410 \\
    \bottomrule
    \end{tabular}
    \caption{Statistical significance between vanilla (control) and interest (treatment) personas. $p$-value is computed through Mann-Whitney U test and adjusted through Holm-Bonferroni method. Effect size is rank-biserial coefficient. Personas with $p$-value less than 0.05 are shaded with grey.}
    \label{table:statistical-significance}
\end{table}

\begin{table}[h]
    \centering
    \small
    \begin{tabular}{>{\bfseries}l?{0.3mm}l}
    \toprule
    \textbf{Persona}            & \textbf{Advertised products} \\
    \midrule
    Health \& Fitness           & \cellcolor[HTML]{8adba0} Dehumidifier, Essential oils \\
    \midrule
    Smart Home                  & \cellcolor[HTML]{8adba0} Vacuum cleaner, Vac. clean. accessories  \\
    \midrule
    Religion \& Spirituality    & \cellcolor[HTML]{fcd14f} Wifi router, Kindle, Swarovski  \\
    \midrule
    Pets \& Animals             & \cellcolor[HTML]{fcd14f} PC files copying/switching software  \\
    \bottomrule
    \end{tabular}
    \caption{Ads from Amazon on interest personas. \colorbox[HTML]{8adba0} {Green} represents unique ads with apparent relevance to the persona. \colorbox[HTML]{fcd14f} {Yellow} represents unique ads that repeat across iterations but do not have any apparent relevance to the persona.} 
    \label{table:ad-content}
\end{table}

\begin{figure}[!t]
    \centering
    \subfloat[Dehumidifier ad in Health \& Fitness]{
        \frame{\includegraphics[width=0.60\columnwidth,trim=7cm 0.5cm 3.5cm 0.4cm, clip]{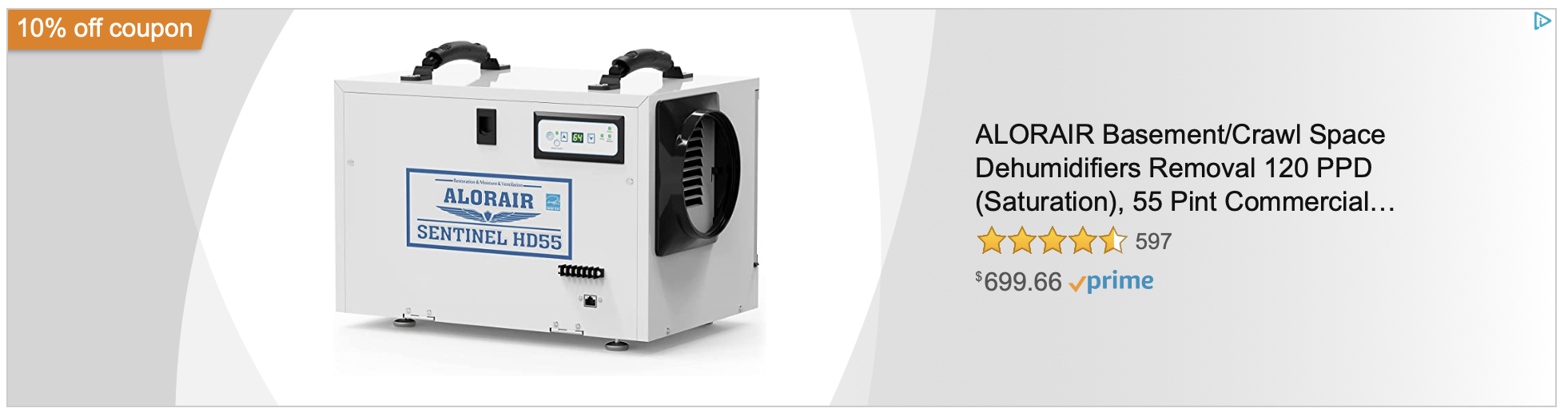}}
        \label{figure:unique-ad-relevance}
    }
    \subfloat[Essential oils ad in Health \& Fitness]{
        \frame{\includegraphics[width=0.30\columnwidth,trim=0.7cm 1.2cm 0.7cm 1.2cm, clip]{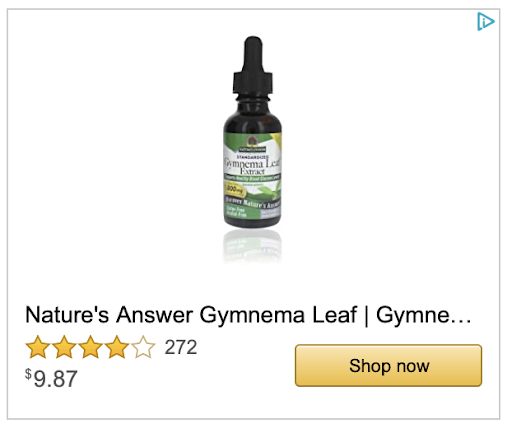}}
        \label{figure:unique-ad-relevance-2}
    }\\
    \subfloat[Vacuum cleaner ad in Smart Home]{
        \frame{\includegraphics[width=0.60\columnwidth,trim=1.2cm 0.3cm 1cm 0.4cm, clip]{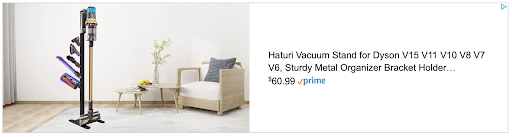}}
        \label{figure:unique-ad-relevance-3}
    }
    \subfloat[Eero WiFi ad in Religion \& Spirituality]{
        \frame{\includegraphics[width=0.30\columnwidth,trim=0.1cm 0.3cm 0.3cm 0.4cm, clip]{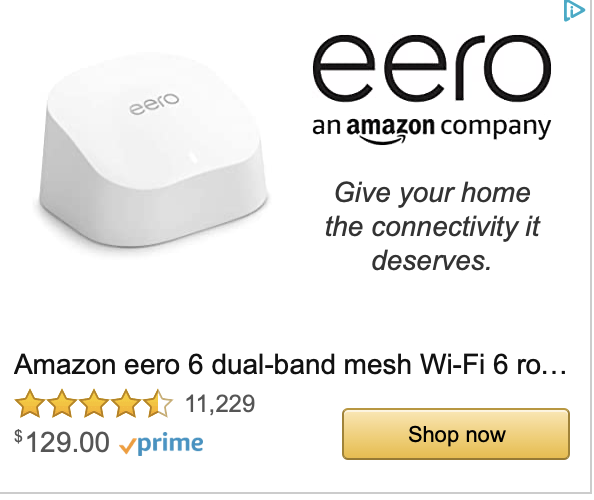}}
        \label{figure:unique-ad-no-relevance}
    }
    \caption{Unique and repeated ads in interest personas.}
    \label{figure:personalized-ads}
\end{figure}

Ads from Amazon do seem to be personalized to personas.
Table \ref{table:ad-content} presents the unique ads from Amazon that show apparent personalization.
Health \& Fitness and Smart Home personas receive unique ads with apparent personalization, whereas Religion \& Spirituality and Pets \& Animals receive unique ads but without any apparent personalization.
The dehumidifier ad (Figure \ref{figure:unique-ad-relevance}) appears to have an association with the \textit{Air Quality Report} skill \cite{air_quality_report_skill} and the essential oils ad (Figure \ref{figure:unique-ad-relevance-2}) appears to have an association with the \textit{Essential Oil Benefits} skill \cite{Essential_Oil_Benefits_skill} in Health \& Fitness persona. 
The dehumidifier ad appeared 7 times across 5 iterations and the essential oils ad appeared once in Health \& Fitness persona. 
The vacuum cleaner and vacuum cleaner accessories ads (Figure \ref{figure:unique-ad-relevance-3}) from Dyson appear to have an association with the \textit{Dyson} skill \cite{dyson_skill}; both ads appeared once in Smart Home persona.
We notice several ads repeated across iterations in Religion \& Spirituality and Pets \& Animals personas that do not seem to have any apparent personalization. 
For example, Amazon Eero WiFi (Figure \ref{figure:unique-ad-no-relevance}), Amazon Kindle, and Swarovski ads exclusively appeared on 12, 14, 2 times across 8, 4, and 2 iterations, respectively in Religion \& Spirituality persona. 
Similarly, PC files copying/switching software ad appeared 4 times in 2 iterations in Pets \& Animals persona.

We also manually analyze ads from installed skill vendors and Amazon with only skill installation. 
We receive a total of 35 and 117 ads from skill vendors and Amazon with only skill installation, respectively.
Out of the 35 ads from installed skills vendors, 16, 9, 2, and 2 are from Microsoft, SimpliSafe, Ring, and SharkNinja in Smart Home persona, respectively. 
Samsung, LG, and ATT also serve one ad each in Smart Home persona.
Honda and Dodge serve one ad each in Connected Car and one Starbucks ad appeared in Wine \& Beverages.
However, none of the ads from installed skills vendors are exclusive to the personas where their skills are installed, which indicates that these ads do not reveal obvious personalization. 
In case of Amazon, out of 117 ads, only two ads are unique to Health \& Fitness persona, i.e., an ad for an electric toothbrush appearing once and an ad for an air fryer toaster appearing 4 times. 
However, similar to the ads from skill vendors, ads from Amazon also lack an apparent relevance to the personas as per our rubric, i.e., Health \& Fitness persona does not have any skills related to electric toothbrush or an air fryer toaster oven.

Since we do not find a strong targeting signal in personas with only skill installation, we do not further analyze this case.

\subsection{Sharing beyond the observed endpoints}
\label{sec:data-sharing-analysis}
Next, we infer the potential sharing of smart speaker interaction metadata from Amazon and third party skills, with other online services, not necessarily observable from the smart speaker.

\subsubsection{Some advertisers sync their cookies with Amazon and bid higher than non-cookie syncing advertisers.}
\label{sub-sec:cookie-syncing}
To target personalized ads, advertisers share user data with each other. 
Typically, unique user identifiers, e.g., cookies, are shared at the client side with cookie syncing and user interest data is synced at the server side \cite{Bashir16TrackingFlowsUsenix}. 
We analyze cookie syncing instances that involve Amazon advertising services in the web traffic captured while collecting ads (Section \ref{subsection:capturing-advertisements}).
We note that 41 third parties sync their cookies with Amazon across all Echo interest personas.
Amazon did not sync its cookies with any advertiser.\footnote{We analyze the OpenWPM datasets released by prior work \cite{Iqbal22USENIXKhaleesi} to validate that Amazon's cookie syncing behavior is not unique to our dataset.} 
The one sided cookie-syncs could be explained by Amazon advertising's recent services for central identity resolution \cite{amazon_identity}.

\begin{figure}[t]
    \centering
    \includegraphics[width=\columnwidth]{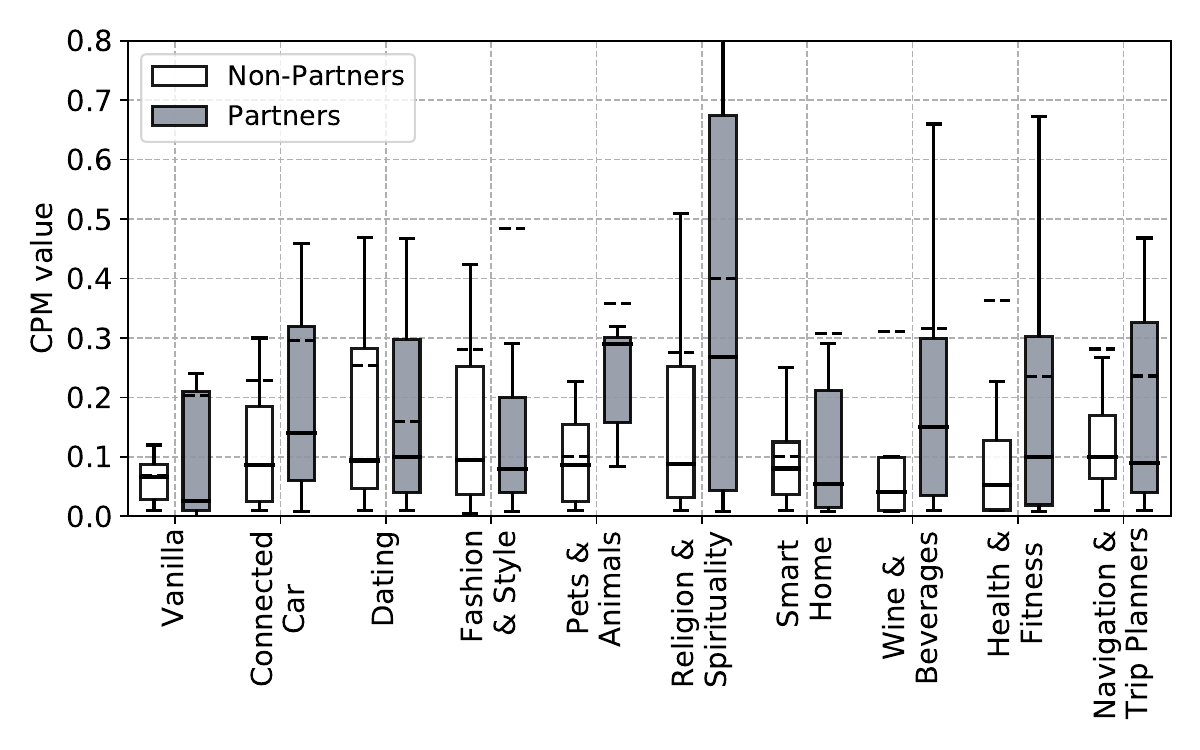}
    \caption{Bid values across personas on common ad slots distributed by Amazon's advertising partners.}
    \label{figure:partner-non-partner}
\end{figure}

To infer potential data sharing by Amazon, we compare and contrast the bid values by Amazon's partners (i.e., cookie syncing advertisers) and non-partner advertisers.
Figure \ref{figure:partner-non-partner} presents the bid values on common ad slots by Amazon's partner and non-partner advertisers.
We note that the bids by partner advertisers are higher than that by non-partner advertisers on most personas.
Table \ref{table:bids-median-mean-partner-non-partner} shows the median and mean bid values by partner and non-partner advertisers.
It can be seen from the table that both median and mean bid values from partners are high for 6 and 7 personas, respectively, as compared to bids from non-partners.
Median bid values are as much as 3$\times$ higher for Pets \& Animals, Religion \& Spirituality, and Wine \& Beverages personas, while mean bid values are 3$\times$ higher for Pets \& Animals, Smart Home, and vanilla personas.
It is noteworthy that Amazon's advertising partners further sync their cookies with 247 other third parties, including advertising services.
Such cookie syncs may lead to the propagation of user data in the advertising ecosystem.

\begin{table}[t]
    \centering
    \small
    \begin{tabular}{>{\bfseries}l?{0.3mm}cccc}
    \toprule
                          & \multicolumn{2}{c}{\textbf{Partner}} & \multicolumn{2}{c}{\textbf{Non-partner}} \\
    \textbf{Persona}                     & \textbf{Median} & \textbf{Mean} & \textbf{Median} & \textbf{Mean} \\
    \midrule
    Connected Car               &  0.140    & 0.296     &   0.086    &  0.228  \\
    Dating                      &  0.099    & 0.159     &   0.094    &  0.254  \\
    Fashion \& Style            &  0.080    & 0.485     &   0.095    &  0.281 \\
    Pets \& Animals             &  0.290    & 0.358     &   0.087    &  0.101  \\
    Religion \& Spirituality    &  0.268    & 0.400     &   0.088    &  0.276   \\
    Smart Home                  &  0.054    & 0.307     &   0.080    &  0.101  \\
    Wine \& Beverages           &  0.150    & 0.316     &   0.041    &  0.310  \\
    Health \& Fitness           &  0.099    & 0.235     &   0.053    &  0.363  \\
    Navigation \& Trip Plan. & 0.090     & 0.236     &   0.100    &  0.281 \\
    Vanilla                     &  0.025    & 0.203  & 0.352  &   0.066 \\
    \bottomrule
    \end{tabular}
    \caption{Median and mean bid values for personas from Amazon's partner and non-partner advertisers.}
    \label{table:bids-median-mean-partner-non-partner}
\end{table}

\subsubsection{It is unclear whether skills play a role in targeting of personalized ads.}
We now discuss Amazon's and skills' role in higher bids and targeting of personalized ads.
Since all interactions are mediated through Amazon, Amazon has the best vantage point to infer personas' interests and target personalized ads.
Specifically, all user commands are interpreted by Amazon and most network requests are routed to/through Amazon (Table \ref{table:domains_list} and Figure \ref{fig:sankey-bigger-version}).
Further, when a persona is logged into its Amazon account, Amazon can access its cookies during web visits. 
In fact, Sections \ref{sub-sec:interest-inference-amazon} and \ref{sub-sec:targeted-ads} already show that Amazon infers users' advertising interests from the metadata of their interaction with Echos and uses the inferred interests to target personalized ads to users.
We also note that Smart Home, Wine \& Beverages, and Navigation \& Trip Planners, personas do not contact any non-Amazon services but still receive high bid values, as compared to vanilla persona.
Amazon further infers discernible interests for the Smart Home  and Fashion \& Style personas (Table \ref{table:amazon-interests}).
These results suggest that Amazon plays a crucial, if not a sole, role in higher bids and targeting of personalized ads.

In contrast, skills can only rely on persona's email address (if given permission), IP address (if skills contact non-Amazon web services directly), and by collaborating with Amazon, to reach to personas.  
Though we allow skills to access email address, we do not log in to any online services (except for Amazon), thus skills cannot use email addresses to target personalized ads. 
Skills that contact non-Amazon web services and skills that collaborate with Amazon can still target ads to users. 
However, we note that only a handful of skills contact few advertising and tracking services (Table \ref{table:domains_list} and Figure \ref{fig:sankey-bigger-version}). 
Similarly, we note that none of the skills re-target ads to personas (Section \ref{sub-sec:targeted-ads}), which implies that Amazon might not be engaging in data sharing partnerships with skills. 

\section{Companies' Representations}
\label{sec:privacy-policy-analysis}
\label{sec:capturing-privacy-policies}

Our auditing framework, so far, measured, directly or indirectly, the actual practices of Amazon as well as skills, with respect to data collection, usage, and sharing.
Companies also make representations and public disclosures, which should accurately and fully disclose their practices.
Such disclosures include press releases, statements on the website, and the -- legally binding -- privacy policies.
In Section \ref{sec:policies}, we focus on and analyze the consistency between the data collection practices of skill vendors (as directly observed in the network traffic) with the statements made in their privacy policies. 

\subsection{Privacy policy analysis}
\label{sec:policies}

\subsubsection{Collecting privacy policies.}
\label{sec:skills-privacy-policy}
We download skills' privacy policies from the \textit{Developer Privacy Policy} link on the skill installation page. 
Recall from Section~\ref{sec:simulating-skill-personas} that we experiment with 450 skills, i.e., top-50 skills from nine categories.
Among the 450 skills, only 214 (47.6\%) skills provide privacy policy links on their installation pages.
The percentage is higher than the statistics reported by prior work~\cite{Lentzsch2021Hey}, which identified that only 28.5\% of the skills provide a privacy policy link.
We surmise that it could be because we investigate popular skills. 
Unfortunately, only 188 skills out of 214 provide a valid privacy policy link. 
Further, among the 188 obtained privacy policies, 129 do not even mention the word ``Alexa'' or ``Amazon'' in their text.
We manually read many of the privacy policies, and notice that they are mostly generic and apply to several products from the same developer. Thus, they do not seem to be specific to Alexa skills.

\subsubsection{Network traffic flows are often inconsistent with privacy policies.}
\label{sub-sec:network-traffic-vs-privacy-policies}
As in prior work \cite{Andow2020Actions, Trimananda2022Ovrseen, Lentzsch2021Hey}, we use on PoliCheck \cite{Andow2020Actions} to evaluate the consistency of network traffic flows with privacy policies.
PoliCheck extracts \textit{$\langle$data type, entity$\rangle$} tuples from the network traffic and the textual disclosures in the privacy policies, and checks the consistency of the two.
However, such analysis requires access to unencrypted network traffic, which is unavailable in our case (see Section~\ref{sec:capturing-network-traffic}). 
Thus, we adapt PoliCheck to perform the analysis only on the endpoints found in the encrypted traffic collected from the \echo{}.

Specifically, we modify PoliCheck to only validate the consistency of endpoint organizations contacted by skills with their privacy policies. 
We update PoliCheck's entity ontology by inspecting the network traffic and including observed endpoints, which we then map to their organization using the methodology described in Section~\ref{sec:capturing-network-traffic}.
Based on the service offered by the organization, it is assigned one or more categories from \textit{platform provider}, \textit{voice assistant service}, \textit{analytic (tracking) provider}, \textit{advertising network}, and \textit{content provider}.
These categories are derived from PoliCheck's entity ontology and terms found in the privacy policies.
We visit the website of each organization to determine the service offered by it.
\textit{Platform provider} and \textit{voice assistant service} labels are only assigned to Amazon. 
We also update Policheck's consistency disclosure definitions.
Specifically, data flows are referred to as (1) \textit{clear}, when the endpoints are disclosed in the privacy policy using the exact organization name; (2) \textit{vague}, when the endpoint is disclosed using category names or \textit{third party}; and (3) \textit{omitted}, when the endpoint is not disclosed at all. 
We do not use \textit{ambiguous} and \textit{incorrect} disclosures because a contradiction cannot be determined without considering data types. 
We label an endpoint as \textit{no policy} when the skill does not provide a privacy policy.

\begin{table*}[t!]
    \centering
    \small
    \begin{tabular}{c|l|l}
    \toprule
    \textbf{Endpoint organization}   & \textbf{Categories in the ontology} & \textbf{Contacted skills} \\
    \midrule
        
        Amazon Technologies, Inc.
        & \makecell[l]{analytic provider, advertising network,\\ content provider, platform provider,\\voice assistant service}
        & \makecell[l]{
          \greenbox{AAA Road Service},
          \greenbox{Salah Time},
          \greenbox{My Dog},
          \greenbox{My Cat},
          \greenbox{Outfit Check!},\\
          \greenbox{Pet Buddy},
          \greenbox{Rain Storm by Healing FM},
          \greenbox{Single Decade Short Rosary},\\
          \greenbox{Islamic Prayer Times},
          \greenbox{Sonos},
          \yellowbox{136 skills},
          \redbox{42 skills},
          \graybox{258 skills}
        }  \\\hline
    
        Chartable Holding Inc
        & analytic provider, advertising network
        & \yellowbox{Garmin},
          \graybox{Makeup of the Day},
          \graybox{My Tesla (Unofficial)} \\\hline
          
        DataCamp Limited
        & content provider
        & \graybox{Relaxing Sounds: Spa Music},
          \graybox{Comfort My Dog},
          \graybox{Calm My Cat} \\\hline
          
        Dilli Labs LLC
        & content provider
        & \makecell[l]{
          \yellowbox{VCA Animal Hospitals},
          \redbox{EcoSmart Live},
          \graybox{Dog Squeaky Toy},\\
          \graybox{Relax My Pet},
          \graybox{Cat Sounds},
          \graybox{Dinosaur Sounds},
          \graybox{Hush Puppy},\\
          \graybox{Calm My Dog},
          \graybox{Calm My Pet}
        }\\\hline
        
        Garmin International
        & content provider
        & \greenbox{Garmin} \\\hline
        
        Liberated Syndication
        & analytic provider, advertising network
        & \graybox{Calm My Pet}, \gray{Al’s Dog Training Tips} \\\hline
        
        National Public Radio, Inc.
        & content provider
        & \graybox{Makeup of the Day}, \gray{Men's Finest Daily Fashion Tip} \\\hline
        
        Philips International B.V.
        & content provider
        & \graybox{Say a Prayer}, \gray{Angry Girlfriend} \\\hline
        
        Podtrac Inc
        & analytic provider, advertising network
        & \makecell[l]{
          \yellowbox{Garmin},
          \yellowbox{Gwynnie Bee},
          \yellowbox{Genesis},
          \graybox{Men's Finest Daily Fashion Tip},\\
          \graybox{Love Trouble},
          \graybox{Makeup of the Day}, 
          \graybox{Dating \& Relationship Tips}
        } \\\hline
        
        Spotify AB
        & analytic provider, advertising network
        & \makecell[l]{
          \yellowbox{Gwynnie Bee},
          \yellowbox{Genesis},
          \graybox{Dating and Relationship Tips and advices},\\
          \graybox{Makeup of the Day},
          \graybox{Men's Finest Daily Fashion Tip},
          \graybox{Love Trouble}
        } \\\hline
        
        Triton Digital, Inc.
        & analytic provider, advertising network
        & \yellowbox{Garmin}, \yellowbox{Charles Stanley Radio}  \\\hline
        
        Voice Apps LLC
        & content provider
        & \makecell[l]{
          \redbox{Prayer Time},
          \redbox{Charles Stanley Radio},
          \graybox{Morning Bible Inspiration},\\
          \graybox{Holy Rosary},
          \graybox{meal prayer},
          \graybox{Halloween Sounds},
          \graybox{Bible Trivia}
        } \\\hline
        
        Life Covenant Church, Inc.
        & content provider
        & \greenbox{YouVersion Bible}, \redbox{Lords Prayer}\\
    \bottomrule
    \end{tabular}
    \caption{Endpoint organizations observed in the network traffic from skills run on the \echo{}: only 32 skills exhibit non-Amazon endpoints. Skills highlighted in \green{green} use the exact organization name in the statement that discloses data collection and sharing by the endpoint. Skills highlighted in \yellow{yellow} use \textit{third party} or other vague terms. Skills highlighted in \red{red} do not declare the contacted endpoint at all. Skills highlighted in \gray{gray} do not provide a privacy policy.}
    \label{table:network-to-policy-consistency-endpoints}
\end{table*}

\textit{Disclosure of platform-party collection.}
Table~\ref{table:network-to-policy-consistency-endpoints} presents the result of our endpoint analysis. 
The table shows that only 10 privacy policies clearly indicate that personal information is collected by Amazon. 
For example, the \textit{Sonos} skill~\cite{SonosSkill} clearly states that voice recordings are collected by Amazon. 
Furthermore, we find that 136 skills vaguely disclose that their network traffic may go to Amazon. 
For example, the \textit{Harmony} skill~\cite{HarmonySkill} privacy policy mentions sending data but without referring to the name of the entity:
\textit{``Circle products may send pseudonymous information to an analytics tool, including timestamps, transmission statistics, feature usage, performance metrics, errors, etc.''}

\textit{Disclosure of first-party collection.}
We find that 32 skills connect to non platform-party endpoints.
Among them, 10 provide privacy policies and only six have at least one clear or vague disclosure. 
The only two clearly disclosed first-party endpoints are in the privacy policies of the \textit{YouVersion Bible}~\cite{YouVersionBibleSkill} and \textit{Garmin}~\cite{GarminSkill} skills: they correspond to the organizations that are the developers of the skills.

\textit{Disclosure of third party collection.}
Many skills rely on third party organizations, e.g., Liberated Syndication, Podtrac, Spotify and Triton Digital, which provide audio content distribution and tracking/advertising services. 
However, only a few skills disclose data collection and sharing with third party organizations in their privacy policies, and when they do, they use vague terms. 
For example, the \textit{Charles Stanley Radio} skill~\cite{CharlesStanleyRadioSkill} uses the term {\em``external service providers''} to refer to third party organizations in its privacy policy. 
Another example is the \textit{VCA Animal Hospitals} skill that  uses the blanket term {\em``third parties''} to refer to all third party organizations in its privacy policy~\cite{VCAAnimalHospitalSkill}.

\subsection{Alexa Echo data processing policies}
\label{sec:amazon-disclosure}
The Alexa Echo smart speaker data collection practices are clearly stated in Amazon's privacy policy, i.e., Amazon collects data when users ``talk to or otherwise interact with our Alexa Voice service'' \cite{amazon_privacy_policy}. 
However, we did not find similar information about the usage of the Alexa Echo interaction data for user interest inference for ad targeting.  
We also explored Amazon's dedicated Alexa specific policies, i.e., Alexa Privacy Hub \cite{alexa_privacy_hub_archived} and Alexa Device FAQs \cite{alexa_device_faq_archived}, but similar to privacy policy, we did not find any information about the usage of Alexa Echo interaction data for ad targeting, at the time of our research.  
However, after our work's preprint was released and Amazon was made aware~\cite{amazon-response}, Amazon updated the Alexa Privacy Hub \cite{alexa_privacy_hub} and the Alexa Device FAQs \cite{alexa_device_faq} to include that Alexa Echo interaction data is used for ad targeting.

\section{Discussion}
\label{sec:discussion}

\subsection{Parallels with other IoT platforms}
\label{appendix:iot-parallels}

\subsubsection{Related platform-agnostic IoT works.} Several IoT works have measured network traffic to detect data collection and sharing. 
For example,~\cite{mazhar2020characterizing,huang2020iot,ren-imc19, saidi-imc20,mandalari-pets21} have shown that tracking is common in several IoT platforms, regardless of the presence of specific apps/skills. 
In contrast to prior work, our study identifies that Alexa Echo smart speakers contact previously unreported endpoints from Amazon, skills vendors, and third parties.
For example, with respect to the endpoints reported in a 2021 study~\cite{mandalari-pets21}, we have observed 4 new Amazon domains (\url{acsechocaptiveportal.com}, \url{amazon-dss.com}, \url{a2z.com}, \url{amazonalexa.com}), 2 skills-specific endpoints (see \emph{skills} row in Table~\ref{table:domains_list}) and 12 new third party endpoints (see \emph{third party} row in Table~\ref{table:domains_list}).
A possible explanation could be the change in Alexa Echo ecosystem since it was last studied, e.g., \url{api.amazonalexa.com} may have replaced \url{api.amazon.com}, which was no longer contacted.

\subsubsection{Related platform-specific IoT works.} 
Compared to prior work on smart TVs~\cite{varmarken2020tv,mohajeri2019watching} and VR headsets~\cite{Trimananda2022Ovrseen}, we found less observable data-tracking activity on smart speakers.
However, ad targeting on the web, specifically from partner advertisers, indicates that data sharing may still be happening.
A possible explanation could be the server-side interfaces from smart speaker platform that expose data for advertising purposes.

\subsection{Possible mitigations}
\label{appendix:possible-defenses}
\subsubsection{Improved transparency and control for users.} 
Smart speaker users may want to know what data is being collected, how that data is being used, and by whom. 
Our work suggests the need for greater transparency for users about the answer to these questions, as well as better control. 
Such transparency and control might come through a redesign of the platform itself (e.g., improved privacy-related UX, system-level enforcement with information flow control) or through external audits (such as with our framework) and external controls (either technical—e.g., network traffic filtering—and/or policy-based). 
For example, Alexa Echos could be equipped with a debugging interface \cite{echohiddenport}.
Making such an interface available for developers and auditors could provide direct observations of data sharing. 
To limit tracking, a user might use software to selectively block network traffic that is not essential for the skill to work (e.g., using an approach similar to \cite{mandalari-pets21}).

\subsubsection{Limiting user interaction data.}
To limit the sharing of data, one can offload the wake-word detection and transcription functions of the Alexa platform with offline tools such as \cite{porcupine, rhasspy}, and just send to the Alexa platform the transcribed commands using their textual API with no loss of functionality. 
Data sharing to only one vendor could also be limited by allowing users an option to install voice assistants from their preferred vendor, similar to apps on mobile devices.

\subsection{Generalizability to other platforms}
The modularity of our framework makes it suitable to be generalized across other smart speaker platforms and even other IoT platforms, such as smart TVs and AR/VR. 
The core idea behind our framework of exposing data and inferring its usage is universally applicable. 
However, other platforms may require implementation changes to various modules of our framework. 
For example, our data exposure module that exposes data by uttering voice commands may be readily deployable to other smart speaker platforms but not to the AR/VR platforms. 
On the other hand, our data usage inference module that infers the usage of exposed data through online advertising on the web may be readily deployable across other smart speaker and IoT platforms. 
Similarly, our network traffic capturing and our privacy policy modules could also be readily deployed to other platforms. 
We also envision that other platforms might also be able to extend our framework by incorporating platform specific components. 
For example, smart TV platforms have a mature video advertising ecosystem \cite{Hoomanthesis}, which could be leveraged to strengthen the inference of exposed data.

\subsection{Limitations}
\label{subsection:limitations}
Our simulated interactions with Alexa Echo skills are potentially different from how real users would interact with the skills. 
For example, we utter a specific set of predefined commands and do not complete the conversation flows, e.g., by responding to the follow up questions by the skills. 
However, even with partial conversation flows, we are able to establish that smart speaker interaction metadata is used to infer user interests, which is then used for ad targeting. 
We expect that with more complete conversation flows, the extent of tracking, profiling, and ad targeting would potentially increase even further.

In order to be able to attribute the usage of exposed data to entities in the smart speaker ecosystem (i.e., the smart speaker platforms, skill vendors, and third parties), we do not expose data to other online services.  
However, in a realistic setting, users might expose their data (e.g., emails) to other online services, making it challenging to assess the role of individual services in using user's data. 
Future work could explore mechanisms to disentangle the role of individual services when data is exposed to several online services.

We currently visually inspect and analyze ad content, which can lead to subjective assessment and also hinder reproducibility. 
Future work could address that limitation by automating ad content analysis (e.g., with the help of machine learning) or by leveraging crowdsourcing techniques used by prior research for ad content analysis~\cite{zeng2021makes,Bashir16TrackingFlowsUsenix} to foster reproducibility.

\subsection{Ethics \& Disclosure}
\label{appendix:ethics}
We visit websites to collect ads and their associated bids, which results in ad impressions and could potentially cause advertisers to lose some revenue.
However, we only visit the minimal number of websites necessary to establish statistical confidence in our inferences, to limit the economic impact of our experiments. 
Note that this concern is common to all web measurement studies and we follow commonly accepted practices to minimize the impact of our measurements.

We did not directly disclose our findings to Amazon because traditional vulnerability disclosures assume overlooked issues, e.g., a security bug because of an implementation flaw.
In our study, the issues we identified seem to be part of the design of the (eco)system and the purpose of our study is to bring public transparency.
In fact, after our work's preprint was released and Amazon was made aware~\cite{amazon-response}, Amazon updated its disclosure to include that it uses smart speaker interaction data for ad targeting~\cite{amazon_manage_preferences}.
We have also shared our findings in a public forum at the Federal Trade Commission (FTC)~\cite{ftcprivacycon2022umar}.

\section{Conclusion}
\label{sec:conclusion}
In this paper, we audited data collection, usage, and sharing practices in the Alexa Echo smart speaker ecosystem.
Our results indicated that \echo{} interactions were tracked by both Amazon and third parties.
We also found that Amazon used \echo{} interaction data to infer user interests and then used those inferences for ad targeting, which was not clearly stated in Amazon's policies before our research and public disclosure. 
In many instances, skills did not clearly disclose their data collection practices in their privacy policies, did not provide any privacy policy, or did not reference the platform's privacy policy.
Given these findings, there is a clear need for increased transparency---by using auditing tools such as ours---on the practices of smart speaker platforms and third parties operating on them.
Our auditing framework and results may be useful to several stakeholders, including Amazon and skill developers (for internal privacy audits), policymakers (for crafting and effectively enforcing regulation), and users (as an incentive to guard their privacy using available tools).

We make our code and datasets publicly available at \url{https://privsec-research.github.io/alexaechos}.

\section*{Acknowledgements}
We would like to thank Caelan MacArthur who contributed to the preliminary investigation of this work in summer 2021 as part of the CRA-WP Distributed Research Experiences for Undergraduates (DREU) at the University of California, Davis. 
We would like to thank Jan Sanislo from the University of Washington IT team for their help with setting up a WiFi network for connecting Alexa Echo devices. 
This work is supported in part by the National Science Foundation under awards CNS-1956393, CNS-1955227, CNS-2103439, CNS-2103038, CNS-2138139, CNS-2114230, CNS-1909020, Computing Research Association for the CIFellows 2021 Project under Grant CNS-2127309, the Northeastern University Future Faculty Fellowship (2021), and the Consumer Reports Innovation Lab.

\bibliographystyle{ACM-Reference-Format}
\balance
\bibliography{references}


\begin{thebibliography}{83}


\ifx \showCODEN    \undefined \def \showCODEN     #1{\unskip}     \fi
\ifx \showDOI      \undefined \def \showDOI       #1{#1}\fi
\ifx \showISBNx    \undefined \def \showISBNx     #1{\unskip}     \fi
\ifx \showISBNxiii \undefined \def \showISBNxiii  #1{\unskip}     \fi
\ifx \showISSN     \undefined \def \showISSN      #1{\unskip}     \fi
\ifx \showLCCN     \undefined \def \showLCCN      #1{\unskip}     \fi
\ifx \shownote     \undefined \def \shownote      #1{#1}          \fi
\ifx \showarticletitle \undefined \def \showarticletitle #1{#1}   \fi
\ifx \showURL      \undefined \def \showURL       {\relax}        \fi
\providecommand\bibfield[2]{#2}
\providecommand\bibinfo[2]{#2}
\providecommand\natexlab[1]{#1}
\providecommand\showeprint[2][]{arXiv:#2}

\bibitem[\protect\citeauthoryear{{Aaron M Spelling}}{{Aaron M
  Spelling}}{2022}]%
        {Dating-and-Relationship-Tips-and-advices-skills}
\bibfield{author}{\bibinfo{person}{{Aaron M Spelling}}.}
  \bibinfo{year}{2022}\natexlab{}.
\newblock \bibinfo{title}{{Dating and Relationship Tips and advices}}.
\newblock \bibinfo{howpublished}{\url{https://www.amazon.com/dp/B07YCKFCCF}}.
\newblock


\bibitem[\protect\citeauthoryear{{Amazon.com, Inc.}}{{Amazon.com,
  Inc.}}{2022a}]%
        {alexa_device_faq}
\bibfield{author}{\bibinfo{person}{{Amazon.com, Inc.}}}
  \bibinfo{year}{2022}\natexlab{a}.
\newblock \bibinfo{title}{Alexa and Alexa Device FAQs}.
\newblock
  \bibinfo{howpublished}{\url{https://www.amazon.com/gp/help/customer/display.html?nodeId=201602230}}.
\newblock


\bibitem[\protect\citeauthoryear{{Amazon.com, Inc.}}{{Amazon.com,
  Inc.}}{2022b}]%
        {alexa_device_faq_archived}
\bibfield{author}{\bibinfo{person}{{Amazon.com, Inc.}}}
  \bibinfo{year}{2022}\natexlab{b}.
\newblock \bibinfo{title}{Alexa and Alexa Device FAQs (archived September
  2022)}.
\newblock
  \bibinfo{howpublished}{\url{https://web.archive.org/web/20220901073936/http://www.amazon.com/gp/help/customer/display.html?nodeId=201602230}}.
\newblock


\bibitem[\protect\citeauthoryear{{Amazon.com, Inc.}}{{Amazon.com,
  Inc.}}{2022c}]%
        {amazon_advertising_restrictions}
\bibfield{author}{\bibinfo{person}{{Amazon.com, Inc.}}}
  \bibinfo{year}{2022}\natexlab{c}.
\newblock \bibinfo{title}{Alexa Blogs: Advertising and Alexa}.
\newblock
  \bibinfo{howpublished}{\url{https://developer.amazon.com/blogs/alexa/post/54c3a0f8-5b29-4071-acd7-2b832b860c83/advertising-and-alexa}}.
\newblock


\bibitem[\protect\citeauthoryear{{Amazon.com, Inc.}}{{Amazon.com,
  Inc.}}{2022d}]%
        {alexa_hosted_skills}
\bibfield{author}{\bibinfo{person}{{Amazon.com, Inc.}}}
  \bibinfo{year}{2022}\natexlab{d}.
\newblock \bibinfo{title}{{Alexa-hosted Skills}}.
\newblock
  \bibinfo{howpublished}{\url{https://developer.amazon.com/en-US/docs/alexa/hosted-skills/build-a-skill-end-to-end-using-an-alexa-hosted-skill.html}}.
\newblock


\bibitem[\protect\citeauthoryear{{Amazon.com, Inc.}}{{Amazon.com,
  Inc.}}{2022e}]%
        {alexa_privacy_hub}
\bibfield{author}{\bibinfo{person}{{Amazon.com, Inc.}}}
  \bibinfo{year}{2022}\natexlab{e}.
\newblock \bibinfo{title}{Alexa Privacy Hub}.
\newblock
  \bibinfo{howpublished}{\url{https://www.amazon.com/Alexa-Privacy-Hub/b?ie=UTF8&node=19149155011}}.
\newblock


\bibitem[\protect\citeauthoryear{{Amazon.com, Inc.}}{{Amazon.com,
  Inc.}}{2022f}]%
        {alexa_privacy_hub_archived}
\bibfield{author}{\bibinfo{person}{{Amazon.com, Inc.}}}
  \bibinfo{year}{2022}\natexlab{f}.
\newblock \bibinfo{title}{Alexa Privacy Hub (archived October 2022)}.
\newblock
  \bibinfo{howpublished}{\url{https://web.archive.org/web/20221010091208/http://www.amazon.com/b/?node=19149155011}}.
\newblock


\bibitem[\protect\citeauthoryear{{Amazon.com, Inc.}}{{Amazon.com,
  Inc.}}{2022g}]%
        {amazon_skill_certification_process}
\bibfield{author}{\bibinfo{person}{{Amazon.com, Inc.}}}
  \bibinfo{year}{2022}\natexlab{g}.
\newblock \bibinfo{title}{{Alexa Skill Certification Requirements}}.
\newblock
  \bibinfo{howpublished}{\url{https://developer.amazon.com/en-US/docs/alexa/custom-skills/certification-requirements-for-custom-skills.html}}.
\newblock


\bibitem[\protect\citeauthoryear{{Amazon.com, Inc.}}{{Amazon.com,
  Inc.}}{2022h}]%
        {amazon_skills_policy}
\bibfield{author}{\bibinfo{person}{{Amazon.com, Inc.}}}
  \bibinfo{year}{2022}\natexlab{h}.
\newblock \bibinfo{title}{{Alexa Skills Policy Testing}}.
\newblock
  \bibinfo{howpublished}{\url{https://developer.amazon.com/en-US/docs/alexa/custom-skills/policy-testing-for-an-alexa-skill.html}}.
\newblock


\bibitem[\protect\citeauthoryear{{Amazon.com, Inc.}}{{Amazon.com,
  Inc.}}{2022i}]%
        {amazon_skills_privacy_policy}
\bibfield{author}{\bibinfo{person}{{Amazon.com, Inc.}}}
  \bibinfo{year}{2022}\natexlab{i}.
\newblock \bibinfo{title}{{Alexa Skills Privacy Requirements}}.
\newblock
  \bibinfo{howpublished}{\url{https://developer.amazon.com/en-US/docs/alexa/custom-skills/security-testing-for-an-alexa-skill.html\#25-privacy-requirements}}.
\newblock


\bibitem[\protect\citeauthoryear{{Amazon.com, Inc.}}{{Amazon.com,
  Inc.}}{2022j}]%
        {amazon_request_pi}
\bibfield{author}{\bibinfo{person}{{Amazon.com, Inc.}}}
  \bibinfo{year}{2022}\natexlab{j}.
\newblock \bibinfo{title}{Amazon: Request Your Data}.
\newblock
  \bibinfo{howpublished}{\url{https://www.amazon.com/gp/privacycentral/dsar/preview.html}}.
\newblock


\bibitem[\protect\citeauthoryear{{Amazon.com, Inc.}}{{Amazon.com,
  Inc.}}{2022k}]%
        {amazon_privacy_policy}
\bibfield{author}{\bibinfo{person}{{Amazon.com, Inc.}}}
  \bibinfo{year}{2022}\natexlab{k}.
\newblock \bibinfo{title}{{Amazon.com Privacy Notice}}.
\newblock
  \bibinfo{howpublished}{\url{https://www.amazon.com/gp/help/customer/display.html?nodeId=GX7NJQ4ZB8MHFRNJ}}.
\newblock


\bibitem[\protect\citeauthoryear{{Amazon.com, Inc.}}{{Amazon.com,
  Inc.}}{2022l}]%
        {amazon_audioads_budget}
\bibfield{author}{\bibinfo{person}{{Amazon.com, Inc.}}}
  \bibinfo{year}{2022}\natexlab{l}.
\newblock \bibinfo{title}{{Audio Ads -- Create audio advertising campaigns}}.
\newblock
  \bibinfo{howpublished}{\url{https://advertising.amazon.com/en-ca/solutions/products/audio-ads}}.
\newblock


\bibitem[\protect\citeauthoryear{{Amazon.com, Inc.}}{{Amazon.com,
  Inc.}}{2022m}]%
        {AlexaCertification}
\bibfield{author}{\bibinfo{person}{{Amazon.com, Inc.}}}
  \bibinfo{year}{2022}\natexlab{m}.
\newblock \bibinfo{title}{{AVS Testing and Certification Process}}.
\newblock
  \bibinfo{howpublished}{\url{https://developer.amazon.com/en-US/docs/alexa/alexa-voice-service/product-testing-overview.html}}.
\newblock


\bibitem[\protect\citeauthoryear{{Amazon.com, Inc.}}{{Amazon.com,
  Inc.}}{2022n}]%
        {amazon_skill_permission_model}
\bibfield{author}{\bibinfo{person}{{Amazon.com, Inc.}}}
  \bibinfo{year}{2022}\natexlab{n}.
\newblock \bibinfo{title}{{Configure Permissions for Customer Information in
  Your Skill}}.
\newblock
  \bibinfo{howpublished}{\url{https://developer.amazon.com/en-US/docs/alexa/custom-skills/configure-permissions-for-customer-information-in-your-skill.html}}.
\newblock


\bibitem[\protect\citeauthoryear{{Amazon.com, Inc.}}{{Amazon.com,
  Inc.}}{2022o}]%
        {amazon_manage_preferences}
\bibfield{author}{\bibinfo{person}{{Amazon.com, Inc.}}}
  \bibinfo{year}{2022}\natexlab{o}.
\newblock \bibinfo{title}{{Managing advertising preferences on Alexa}}.
\newblock
  \bibinfo{howpublished}{\url{https://www.amazon.com/b/?node=98592480011}}.
\newblock


\bibitem[\protect\citeauthoryear{{Amazon.com, Inc.}}{{Amazon.com,
  Inc.}}{2022p}]%
        {alexa_skill_design_docs}
\bibfield{author}{\bibinfo{person}{{Amazon.com, Inc.}}}
  \bibinfo{year}{2022}\natexlab{p}.
\newblock \bibinfo{title}{Module 2: Design an Engaging Voice User Interface}.
\newblock
  \bibinfo{howpublished}{\url{https://developer.amazon.com/en-US/alexa/alexa-skills-kit/get-deeper/tutorials-code-samples/build-an-engaging-alexa-skill/module-2}}.
\newblock


\bibitem[\protect\citeauthoryear{{Amazon.com, Inc.}}{{Amazon.com,
  Inc.}}{2022q}]%
        {amazon_advertising_restrictions_2}
\bibfield{author}{\bibinfo{person}{{Amazon.com, Inc.}}}
  \bibinfo{year}{2022}\natexlab{q}.
\newblock \bibinfo{title}{Policy Testing}.
\newblock
  \bibinfo{howpublished}{\url{https://developer.amazon.com/en-US/docs/alexa/custom-skills/policy-testing-for-an-alexa-skill.html\#advertising}}.
\newblock


\bibitem[\protect\citeauthoryear{Andow, Mahmud, Whitaker, Enck, Reaves, Singh,
  and Egelman}{Andow et~al\mbox{.}}{2020}]%
        {Andow2020Actions}
\bibfield{author}{\bibinfo{person}{Benjamin Andow},
  \bibinfo{person}{Samin~Yaseer Mahmud}, \bibinfo{person}{Justin Whitaker},
  \bibinfo{person}{William Enck}, \bibinfo{person}{Bradley Reaves},
  \bibinfo{person}{Kapil Singh}, {and} \bibinfo{person}{Serge Egelman}.}
  \bibinfo{year}{2020}\natexlab{}.
\newblock \showarticletitle{Actions Speak Louder than Words: {Entity-Sensitive}
  Privacy Policy and Data Flow Analysis with {PoliCheck}}. In
  \bibinfo{booktitle}{\emph{29th USENIX Security Symposium (USENIX Security
  20)}}. \bibinfo{publisher}{USENIX Association}, \bibinfo{address}{Boston},
  \bibinfo{pages}{985--1002}.
\newblock
\showISBNx{978-1-939133-17-5}
\urldef\tempurl%
\url{https://www.usenix.org/conference/usenixsecurity20/presentation/andow}
\showURL{%
\tempurl}


\bibitem[\protect\citeauthoryear{Barceló-Armada, Castell-Uroz, and
  Barlet-Ros}{Barceló-Armada et~al\mbox{.}}{2022}]%
        {BARCELOARMADA2022108782}
\bibfield{author}{\bibinfo{person}{Rubén Barceló-Armada},
  \bibinfo{person}{Ismael Castell-Uroz}, {and} \bibinfo{person}{Pere
  Barlet-Ros}.} \bibinfo{year}{2022}\natexlab{}.
\newblock \showarticletitle{Amazon Alexa traffic traces}.
\newblock \bibinfo{journal}{\emph{Computer Networks}}  \bibinfo{volume}{205}
  (\bibinfo{year}{2022}), \bibinfo{pages}{108782}.
\newblock


\bibitem[\protect\citeauthoryear{Bashir, Arshad, Robertson, and Wilson}{Bashir
  et~al\mbox{.}}{2016}]%
        {Bashir16TrackingFlowsUsenix}
\bibfield{author}{\bibinfo{person}{Muhammad~Ahmad Bashir},
  \bibinfo{person}{Sajjad Arshad}, \bibinfo{person}{William Robertson}, {and}
  \bibinfo{person}{Christo Wilson}.} \bibinfo{year}{2016}\natexlab{}.
\newblock \showarticletitle{Tracing Information Flows Between Ad Exchanges
  Using Retargeted Ads}. In \bibinfo{booktitle}{\emph{25th USENIX Security
  Symposium (USENIX Security 16)}}. \bibinfo{publisher}{USENIX Association},
  \bibinfo{address}{Austin}, \bibinfo{pages}{481--496}.
\newblock


\bibitem[\protect\citeauthoryear{Cheng, Wilson, Liao, Young, Dong, and
  Hu}{Cheng et~al\mbox{.}}{2020}]%
        {Cheng2020SkillPolciyViolationCCS}
\bibfield{author}{\bibinfo{person}{Long Cheng}, \bibinfo{person}{Christin
  Wilson}, \bibinfo{person}{Song Liao}, \bibinfo{person}{Jeffrey Young},
  \bibinfo{person}{Daniel Dong}, {and} \bibinfo{person}{Hongxin Hu}.}
  \bibinfo{year}{2020}\natexlab{}.
\newblock \showarticletitle{Dangerous Skills Got Certified: Measuring the
  Trustworthiness of Skill Certification in Voice Personal Assistant
  Platforms}. In \bibinfo{booktitle}{\emph{Proceedings of the 2020 ACM SIGSAC
  Conference on Computer and Communications Security}}
  \emph{(\bibinfo{series}{CCS '20})}. \bibinfo{publisher}{Association for
  Computing Machinery}, \bibinfo{address}{Virtual},
  \bibinfo{pages}{1699–1716}.
\newblock


\bibitem[\protect\citeauthoryear{Claburn}{Claburn}{2022}]%
        {amazon-response}
\bibfield{author}{\bibinfo{person}{Thomas Claburn}.}
  \bibinfo{year}{2022}\natexlab{}.
\newblock \bibinfo{title}{{Study: How Amazon uses Echo smart speaker
  conversations to target ads}}.
\newblock
  \bibinfo{howpublished}{\url{https://theregister.com/2022/04/27/amazon_audio_data/}}.
\newblock


\bibitem[\protect\citeauthoryear{Cook, Nithyanand, and Shafiq}{Cook
  et~al\mbox{.}}{2020}]%
        {Cook20HeaderBiddingPETS}
\bibfield{author}{\bibinfo{person}{John Cook}, \bibinfo{person}{Rishab
  Nithyanand}, {and} \bibinfo{person}{Zubair Shafiq}.}
  \bibinfo{year}{2020}\natexlab{}.
\newblock \showarticletitle{Inferring Tracker-Advertiser Relationships in the
  Online Advertising Ecosystem using Header Bidding}. In
  \bibinfo{booktitle}{\emph{Proceedings on Privacy Enhancing Technologies}},
  Vol.~\bibinfo{volume}{2020 (1)}. \bibinfo{publisher}{Sciendo},
  \bibinfo{address}{Virtual}, \bibinfo{pages}{65--82}.
\newblock


\bibitem[\protect\citeauthoryear{{Crunchbase Inc.}}{{Crunchbase Inc.}}{2022}]%
        {Crunchbase}
\bibfield{author}{\bibinfo{person}{{Crunchbase Inc.}}}
  \bibinfo{year}{2022}\natexlab{}.
\newblock \bibinfo{title}{{Crunchbase}}.
\newblock \bibinfo{howpublished}{\url{https://www.crunchbase.com/}}.
\newblock


\bibitem[\protect\citeauthoryear{Dubois, Kolcun, Mandalari, Paracha, Choffnes,
  and Haddadi}{Dubois et~al\mbox{.}}{2020}]%
        {dubois2020speakersPETS}
\bibfield{author}{\bibinfo{person}{Daniel~J Dubois}, \bibinfo{person}{Roman
  Kolcun}, \bibinfo{person}{Anna~Maria Mandalari},
  \bibinfo{person}{Muhammad~Talha Paracha}, \bibinfo{person}{David Choffnes},
  {and} \bibinfo{person}{Hamed Haddadi}.} \bibinfo{year}{2020}\natexlab{}.
\newblock \showarticletitle{{When Speakers Are All Ears: Characterizing
  Misactivations of IoT Smart Speakers}}. In
  \bibinfo{booktitle}{\emph{Proceedings on Privacy Enhancing Technologies}},
  Vol.~\bibinfo{volume}{2020 (4)}. \bibinfo{publisher}{Sciendo},
  \bibinfo{address}{Virtual}, \bibinfo{pages}{255--276}.
\newblock


\bibitem[\protect\citeauthoryear{{DuckDuckGo}}{{DuckDuckGo}}{2022}]%
        {DDGTrackerRadar}
\bibfield{author}{\bibinfo{person}{{DuckDuckGo}}.}
  \bibinfo{year}{2022}\natexlab{}.
\newblock \bibinfo{title}{{Tracker Radar (list of entities)}}.
\newblock
  \bibinfo{howpublished}{\url{https://github.com/duckduckgo/tracker-radar/tree/main/entities}}.
\newblock


\bibitem[\protect\citeauthoryear{Durette}{Durette}{2022}]%
        {gtts_python}
\bibfield{author}{\bibinfo{person}{Pierre~N. Durette}.}
  \bibinfo{year}{2022}\natexlab{}.
\newblock \bibinfo{title}{{gTTS (Google Text-to-Speech), a Python library and
  CLI tool to interface with Google Translate text-to-speech API}}.
\newblock \bibinfo{howpublished}{\url{https://pypi.org/project/gTTS/}}.
\newblock


\bibitem[\protect\citeauthoryear{{Dyson Limited}}{{Dyson Limited}}{2022}]%
        {dyson_skill}
\bibfield{author}{\bibinfo{person}{{Dyson Limited}}.}
  \bibinfo{year}{2022}\natexlab{}.
\newblock \bibinfo{title}{{Dyson}}.
\newblock \bibinfo{howpublished}{\url{https://www.amazon.com/dp/B06WVN7SHC}}.
\newblock


\bibitem[\protect\citeauthoryear{Edu, Ferrer-Aran, Such, and Suarez-Tangil}{Edu
  et~al\mbox{.}}{2023}]%
        {edu2021skillvet}
\bibfield{author}{\bibinfo{person}{Jide~S. Edu}, \bibinfo{person}{Xavier
  Ferrer-Aran}, \bibinfo{person}{Jose Such}, {and} \bibinfo{person}{Guillermo
  Suarez-Tangil}.} \bibinfo{year}{2023}\natexlab{}.
\newblock \showarticletitle{{SkillVet: Automated Traceability Analysis of
  Amazon Alexa Skills}}.
\newblock \bibinfo{journal}{\emph{IEEE Transactions on Dependable and Secure
  Computing}} \bibinfo{volume}{20}, \bibinfo{number}{1} (\bibinfo{year}{2023}),
  \bibinfo{pages}{161--175}.
\newblock


\bibitem[\protect\citeauthoryear{Englehardt and Narayanan}{Englehardt and
  Narayanan}{2016}]%
        {Englehardt16MillionSiteMeasurementCCS}
\bibfield{author}{\bibinfo{person}{Steven Englehardt} {and}
  \bibinfo{person}{Arvind Narayanan}.} \bibinfo{year}{2016}\natexlab{}.
\newblock \showarticletitle{Online Tracking: A 1-Million-Site Measurement and
  Analysis}. In \bibinfo{booktitle}{\emph{Proceedings of the 2016 ACM SIGSAC
  Conference on Computer and Communications Security}}
  \emph{(\bibinfo{series}{CCS '16})}. \bibinfo{publisher}{Association for
  Computing Machinery}, \bibinfo{address}{Vienna},
  \bibinfo{pages}{1388–1401}.
\newblock


\bibitem[\protect\citeauthoryear{Fowler}{Fowler}{2019}]%
        {smart_speaker_eavesdropping}
\bibfield{author}{\bibinfo{person}{Geoffrey~A. Fowler}.}
  \bibinfo{year}{2019}\natexlab{}.
\newblock \bibinfo{title}{Alexa has been eavesdropping on you this whole time}.
\newblock
  \bibinfo{howpublished}{\url{https://www.washingtonpost.com/technology/2019/05/06/alexa-has-been-eavesdropping-you-this-whole-time/}}.
\newblock


\bibitem[\protect\citeauthoryear{{Garmin International}}{{Garmin
  International}}{2022}]%
        {GarminSkill}
\bibfield{author}{\bibinfo{person}{{Garmin International}}.}
  \bibinfo{year}{2022}\natexlab{}.
\newblock \bibinfo{title}{{Garmin}}.
\newblock \bibinfo{howpublished}{\url{https://www.amazon.com/dp/B075TRB4V5}}.
\newblock


\bibitem[\protect\citeauthoryear{{Gary Horcher}}{{Gary Horcher}}{2018}]%
        {alexa_misactivation_real_world}
\bibfield{author}{\bibinfo{person}{{Gary Horcher}}.}
  \bibinfo{year}{2018}\natexlab{}.
\newblock \bibinfo{title}{Woman says her Amazon device recorded private
  conversation, sent it out to random contact}.
\newblock
  \bibinfo{howpublished}{\url{https://www.kiro7.com/news/local/woman-says-her-amazon-device-recorded-private-conversation-sent-it-out-to-random-contact/755507974/}}.
\newblock


\bibitem[\protect\citeauthoryear{{Genesis Motors USA}}{{Genesis Motors
  USA}}{2022}]%
        {genesisskill}
\bibfield{author}{\bibinfo{person}{{Genesis Motors USA}}.}
  \bibinfo{year}{2022}\natexlab{}.
\newblock \bibinfo{title}{{Genesis}}.
\newblock \bibinfo{howpublished}{\url{https://www.amazon.com/dp/B01JXP09PI}}.
\newblock


\bibitem[\protect\citeauthoryear{Girish, Hu, Prakash, Dubois, Matic, Yuxing,
  Egelman, Reardon, Tapiador, Choffnes, and Vallina-Rodriguez}{Girish
  et~al\mbox{.}}{2023}]%
        {girish-imc23}
\bibfield{author}{\bibinfo{person}{Aniketh Girish}, \bibinfo{person}{Tianrui
  Hu}, \bibinfo{person}{Vijay Prakash}, \bibinfo{person}{Daniel~J. Dubois},
  \bibinfo{person}{Srdjan Matic}, \bibinfo{person}{Danny Yuxing},
  \bibinfo{person}{Serge Egelman}, \bibinfo{person}{Joel Reardon},
  \bibinfo{person}{Juan Tapiador}, \bibinfo{person}{David Choffnes}, {and}
  \bibinfo{person}{Narseo Vallina-Rodriguez}.} \bibinfo{year}{2023}\natexlab{}.
\newblock \showarticletitle{{In the Room Where It Happens: Characterizing Local
  Communication and Threats in Smart Homes}}. In
  \bibinfo{booktitle}{\emph{Proceedings of the ACM Internet Measurement
  Conference}} \emph{(\bibinfo{series}{IMC '23})}.
  \bibinfo{publisher}{Association for Computing Machinery},
  \bibinfo{address}{Montréal}.
\newblock


\bibitem[\protect\citeauthoryear{{Google, Inc.}}{{Google, Inc.}}{2022a}]%
        {HB_protocol}
\bibfield{author}{\bibinfo{person}{{Google, Inc.}}}
  \bibinfo{year}{2022}\natexlab{a}.
\newblock \bibinfo{title}{{Header Bidding}}.
\newblock
  \bibinfo{howpublished}{\url{https://admanager.google.com/home/resources/feature-brief-open-bidding/}}.
\newblock


\bibitem[\protect\citeauthoryear{{Google, Inc.}}{{Google, Inc.}}{2022b}]%
        {RTB_protocol}
\bibfield{author}{\bibinfo{person}{{Google, Inc.}}}
  \bibinfo{year}{2022}\natexlab{b}.
\newblock \bibinfo{title}{{Real-time Bidding}}.
\newblock
  \bibinfo{howpublished}{\url{https://developers.google.com/authorized-buyers/rtb/start}}.
\newblock


\bibitem[\protect\citeauthoryear{{Google, Inc.}}{{Google, Inc.}}{2022c}]%
        {google_rtb_docs}
\bibfield{author}{\bibinfo{person}{{Google, Inc.}}}
  \bibinfo{year}{2022}\natexlab{c}.
\newblock \bibinfo{title}{{RTB - Cookie Matching}}.
\newblock
  \bibinfo{howpublished}{\url{https://developers.google.com/authorized-buyers/rtb/cookie-guide}}.
\newblock


\bibitem[\protect\citeauthoryear{Hu, Dubois, and Choffnes}{Hu
  et~al\mbox{.}}{2023}]%
        {hu-imc23}
\bibfield{author}{\bibinfo{person}{Tianrui Hu}, \bibinfo{person}{Daniel~J.
  Dubois}, {and} \bibinfo{person}{David Choffnes}.}
  \bibinfo{year}{2023}\natexlab{}.
\newblock \showarticletitle{{BehavIoT: Measuring Smart Home IoT Behavior Using
  Network-Inferred Behavior Models}}. In \bibinfo{booktitle}{\emph{Proceedings
  of the ACM Internet Measurement Conference}} \emph{(\bibinfo{series}{IMC
  '23})}. \bibinfo{publisher}{Association for Computing Machinery},
  \bibinfo{address}{Montréal}.
\newblock


\bibitem[\protect\citeauthoryear{Huang, Apthorpe, Li, Acar, and Feamster}{Huang
  et~al\mbox{.}}{2020}]%
        {huang2020iot}
\bibfield{author}{\bibinfo{person}{Danny~Yuxing Huang}, \bibinfo{person}{Noah
  Apthorpe}, \bibinfo{person}{Frank Li}, \bibinfo{person}{Gunes Acar}, {and}
  \bibinfo{person}{Nick Feamster}.} \bibinfo{year}{2020}\natexlab{}.
\newblock \showarticletitle{IoT Inspector: Crowdsourcing Labeled Network
  Traffic from Smart Home Devices at Scale}. In
  \bibinfo{booktitle}{\emph{Proceedings of the ACM on Interactive, Mobile,
  Wearable and Ubiquitous Technologies}}, Vol.~\bibinfo{volume}{4}.
  \bibinfo{publisher}{Association for Computing Machinery},
  \bibinfo{address}{New York, NY, USA}, Article \bibinfo{articleno}{46},
  \bibinfo{numpages}{21}~pages.
\newblock
\urldef\tempurl%
\url{https://doi.org/10.1145/3397333}
\showDOI{\tempurl}


\bibitem[\protect\citeauthoryear{{ICM}}{{ICM}}{2022}]%
        {air_quality_report_skill}
\bibfield{author}{\bibinfo{person}{{ICM}}.} \bibinfo{year}{2022}\natexlab{}.
\newblock \bibinfo{title}{{Air Quality Report}}.
\newblock \bibinfo{howpublished}{\url{https://www.amazon.com/dp/B01EOFCHMA}}.
\newblock


\bibitem[\protect\citeauthoryear{{In Touch Ministries}}{{In Touch
  Ministries}}{2022}]%
        {CharlesStanleyRadioSkill}
\bibfield{author}{\bibinfo{person}{{In Touch Ministries}}.}
  \bibinfo{year}{2022}\natexlab{}.
\newblock \bibinfo{title}{{Charles Stanley Radio}}.
\newblock \bibinfo{howpublished}{\url{https://www.amazon.com/dp/B07FF2QGXW}}.
\newblock


\bibitem[\protect\citeauthoryear{Iqbal}{Iqbal}{2022}]%
        {ftcprivacycon2022umar}
\bibfield{author}{\bibinfo{person}{Umar Iqbal}.}
  \bibinfo{year}{2022}\natexlab{}.
\newblock \bibinfo{title}{{Your Echos are Heard: Tracking, Profiling, and Ad
  Targeting in the Amazon Smart Speaker Ecosystem, FTC PrivacyCon 2022}}.
\newblock
  \bibinfo{howpublished}{\url{https://www.ftc.gov/news-events/events/2022/11/privacycon-2022}}.
\newblock


\bibitem[\protect\citeauthoryear{Iqbal, Wolfe, Nguyen, Englehardt, and
  Shafiq}{Iqbal et~al\mbox{.}}{2022}]%
        {Iqbal22USENIXKhaleesi}
\bibfield{author}{\bibinfo{person}{Umar Iqbal}, \bibinfo{person}{Charlie
  Wolfe}, \bibinfo{person}{Charles Nguyen}, \bibinfo{person}{Steven
  Englehardt}, {and} \bibinfo{person}{Zubair Shafiq}.}
  \bibinfo{year}{2022}\natexlab{}.
\newblock \showarticletitle{Khaleesi: Breaker of Advertising and Tracking
  Request Chains}. In \bibinfo{booktitle}{\emph{31st USENIX Security
  Symposium}}. \bibinfo{publisher}{USENIX Association},
  \bibinfo{address}{Boston}, \bibinfo{pages}{2911--2928}.
\newblock


\bibitem[\protect\citeauthoryear{{iRobot}}{{iRobot}}{2022}]%
        {irobot_skill}
\bibfield{author}{\bibinfo{person}{{iRobot}}.} \bibinfo{year}{2022}\natexlab{}.
\newblock \bibinfo{title}{{iRobot Home}}.
\newblock \bibinfo{howpublished}{\url{https://www.amazon.com/dp/B06Y3PSHQ3}}.
\newblock


\bibitem[\protect\citeauthoryear{Jin and Wang}{Jin and Wang}{2018}]%
        {jin2018voice-Amazon-Patent}
\bibfield{author}{\bibinfo{person}{Huafeng Jin} {and} \bibinfo{person}{Shuo
  Wang}.} \bibinfo{year}{2018}\natexlab{}.
\newblock \bibinfo{title}{Voice-based determination of physical and emotional
  characteristics of users}.
\newblock
\newblock
\newblock
\shownote{US Patent 10096319B1.}


\bibitem[\protect\citeauthoryear{{Kevel}}{{Kevel}}{2022}]%
        {HB_usage}
\bibfield{author}{\bibinfo{person}{{Kevel}}.} \bibinfo{year}{2022}\natexlab{}.
\newblock \bibinfo{title}{{Header Bidding (HBIX) 2021 Tracker}}.
\newblock \bibinfo{howpublished}{\url{https://www.kevel.co/hbix/}}.
\newblock


\bibitem[\protect\citeauthoryear{Le~Pochat, Van~Goethem, Tajalizadehkhoob,
  Korczy{\'n}ski, and Joosen}{Le~Pochat et~al\mbox{.}}{2019}]%
        {le2019tranco}
\bibfield{author}{\bibinfo{person}{Victor Le~Pochat}, \bibinfo{person}{Tom
  Van~Goethem}, \bibinfo{person}{Samaneh Tajalizadehkhoob},
  \bibinfo{person}{Maciej Korczy{\'n}ski}, {and} \bibinfo{person}{Wouter
  Joosen}.} \bibinfo{year}{2019}\natexlab{}.
\newblock \showarticletitle{Tranco: A Research-Oriented Top Sites Ranking
  Hardened Against Manipulation}. In \bibinfo{booktitle}{\emph{26th Annual
  Network and Distributed System Security Symposium}}.
  \bibinfo{publisher}{Internet Society}, \bibinfo{address}{San Diego}.
\newblock


\bibitem[\protect\citeauthoryear{Lentzsch, Shah, Andow, Degeling, Das, and
  Enck}{Lentzsch et~al\mbox{.}}{2021}]%
        {Lentzsch2021Hey}
\bibfield{author}{\bibinfo{person}{Christopher Lentzsch},
  \bibinfo{person}{Sheel~Jayesh Shah}, \bibinfo{person}{Benjamin Andow},
  \bibinfo{person}{Martin Degeling}, \bibinfo{person}{Anupam Das}, {and}
  \bibinfo{person}{William Enck}.} \bibinfo{year}{2021}\natexlab{}.
\newblock \showarticletitle{Hey Alexa, is this skill safe?: Taking a closer
  look at the Alexa skill ecosystem}. In \bibinfo{booktitle}{\emph{28th Annual
  Network and Distributed System Security Symposium}}. \bibinfo{publisher}{The
  Internet Society}, \bibinfo{address}{San Diego}.
\newblock


\bibitem[\protect\citeauthoryear{{Logitech}}{{Logitech}}{2022}]%
        {HarmonySkill}
\bibfield{author}{\bibinfo{person}{{Logitech}}.}
  \bibinfo{year}{2022}\natexlab{}.
\newblock \bibinfo{title}{{Harmony}}.
\newblock \bibinfo{howpublished}{\url{https://www.amazon.com/dp/B01M4LDPX3}}.
\newblock


\bibitem[\protect\citeauthoryear{Maheshwari}{Maheshwari}{2018}]%
        {nyt_alexa_ads}
\bibfield{author}{\bibinfo{person}{Sapna Maheshwari}.}
  \bibinfo{year}{2018}\natexlab{}.
\newblock \bibinfo{title}{{Hey, Alexa, What Can You Hear? And What Will You Do
  With It?}}
\newblock
  \bibinfo{howpublished}{\url{https://www.nytimes.com/2018/03/31/business/media/amazon-google-privacy-digital-assistants.html}}.
\newblock


\bibitem[\protect\citeauthoryear{Mandalari, Dubois, Kolcun, Paracha, Haddadi,
  and Choffnes}{Mandalari et~al\mbox{.}}{2021}]%
        {mandalari-pets21}
\bibfield{author}{\bibinfo{person}{Anna~Maria Mandalari},
  \bibinfo{person}{Daniel~J Dubois}, \bibinfo{person}{Roman Kolcun},
  \bibinfo{person}{Muhammad~Talha Paracha}, \bibinfo{person}{Hamed Haddadi},
  {and} \bibinfo{person}{David Choffnes}.} \bibinfo{year}{2021}\natexlab{}.
\newblock \showarticletitle{{Blocking Without Breaking: Identification and
  Mitigation of Non-Essential IoT Traffic}}. In
  \bibinfo{booktitle}{\emph{Proceedings on Privacy Enhancing Technologies}},
  Vol.~\bibinfo{volume}{2021 (4)}. \bibinfo{publisher}{Sciendo},
  \bibinfo{address}{Virtual}, \bibinfo{pages}{369--388}.
\newblock


\bibitem[\protect\citeauthoryear{Mazhar and Shafiq}{Mazhar and Shafiq}{2020}]%
        {mazhar2020characterizing}
\bibfield{author}{\bibinfo{person}{M.~Hammad Mazhar} {and}
  \bibinfo{person}{Zubair Shafiq}.} \bibinfo{year}{2020}\natexlab{}.
\newblock \showarticletitle{Characterizing Smart Home IoT Traffic in the Wild}.
  In \bibinfo{booktitle}{\emph{2020 IEEE/ACM Fifth International Conference on
  Internet-of-Things Design and Implementation (IoTDI)}}.
  \bibinfo{publisher}{IEEE}, \bibinfo{address}{Sydney},
  \bibinfo{pages}{203--215}.
\newblock


\bibitem[\protect\citeauthoryear{{Men's Finest}}{{Men's Finest}}{2022}]%
        {mens-finest-daily-fashion-tip-skill}
\bibfield{author}{\bibinfo{person}{{Men's Finest}}.}
  \bibinfo{year}{2022}\natexlab{}.
\newblock \bibinfo{title}{{Men's Finest Daily Fashion Tip}}.
\newblock \bibinfo{howpublished}{\url{https://www.amazon.com/dp/B07CB3ZN6N}}.
\newblock


\bibitem[\protect\citeauthoryear{Moghaddam}{Moghaddam}{2022}]%
        {Hoomanthesis}
\bibfield{author}{\bibinfo{person}{Hooman~Mohajeri Moghaddam}.}
  \bibinfo{year}{2022}\natexlab{}.
\newblock \bibinfo{title}{Tracking and Behavioral Targeting on Connected TV
  Platforms}.  (\bibinfo{year}{2022}).
\newblock
\urldef\tempurl%
\url{https://dataspace.princeton.edu/handle/88435/dsp010p096b14c}
\showURL{%
\tempurl}
\newblock
\shownote{Doctoral Disseration.}


\bibitem[\protect\citeauthoryear{Mohajeri~Moghaddam, Acar, Burgess, Mathur,
  Huang, Feamster, Felten, Mittal, and Narayanan}{Mohajeri~Moghaddam
  et~al\mbox{.}}{2019}]%
        {mohajeri2019watching}
\bibfield{author}{\bibinfo{person}{Hooman Mohajeri~Moghaddam},
  \bibinfo{person}{Gunes Acar}, \bibinfo{person}{Ben Burgess},
  \bibinfo{person}{Arunesh Mathur}, \bibinfo{person}{Danny~Yuxing Huang},
  \bibinfo{person}{Nick Feamster}, \bibinfo{person}{Edward~W. Felten},
  \bibinfo{person}{Prateek Mittal}, {and} \bibinfo{person}{Arvind Narayanan}.}
  \bibinfo{year}{2019}\natexlab{}.
\newblock \showarticletitle{Watching You Watch: The Tracking Ecosystem of
  Over-the-Top TV Streaming Devices}. In \bibinfo{booktitle}{\emph{Proceedings
  of the 2019 ACM SIGSAC Conference on Computer and Communications Security}}
  \emph{(\bibinfo{series}{CCS '19})}. \bibinfo{publisher}{Association for
  Computing Machinery}, \bibinfo{address}{London}, \bibinfo{pages}{131–147}.
\newblock


\bibitem[\protect\citeauthoryear{Nardi}{Nardi}{2019}]%
        {echohiddenport}
\bibfield{author}{\bibinfo{person}{Tom Nardi}.}
  \bibinfo{year}{2019}\natexlab{}.
\newblock \bibinfo{title}{{Uncovering the Echo Dot's Hidden USB Port}}.
\newblock
  \bibinfo{howpublished}{\url{https://hackaday.com/2019/08/15/uncovering-the-echo-dots-hidden-usb-port/}}.
\newblock


\bibitem[\protect\citeauthoryear{Olejnik, Tran, and Castelluccia}{Olejnik
  et~al\mbox{.}}{2014}]%
        {Olejnik14SellingPrivacyNDSS}
\bibfield{author}{\bibinfo{person}{Lukasz Olejnik}, \bibinfo{person}{Minh-Dung
  Tran}, {and} \bibinfo{person}{Claude Castelluccia}.}
  \bibinfo{year}{2014}\natexlab{}.
\newblock \showarticletitle{Selling Off Privacy at Auction}. In
  \bibinfo{booktitle}{\emph{21st Annual Network and Distributed System Security
  Symposium}}. \bibinfo{publisher}{The Internet Society}, \bibinfo{address}{San
  Diego}.
\newblock


\bibitem[\protect\citeauthoryear{Papadopoulos, Kourtellis, Rodriguez, and
  Laoutaris}{Papadopoulos et~al\mbox{.}}{2017}]%
        {Papadopoulos17IMCYouAreTheProduct}
\bibfield{author}{\bibinfo{person}{Panagiotis Papadopoulos},
  \bibinfo{person}{Nicolas Kourtellis}, \bibinfo{person}{Pablo~Rodriguez
  Rodriguez}, {and} \bibinfo{person}{Nikolaos Laoutaris}.}
  \bibinfo{year}{2017}\natexlab{}.
\newblock \showarticletitle{If You Are Not Paying for It, You Are the Product:
  How Much Do Advertisers Pay to Reach You?}. In
  \bibinfo{booktitle}{\emph{Proceedings of the 2017 Internet Measurement
  Conference}}. \bibinfo{publisher}{Association for Computing Machinery},
  \bibinfo{address}{London}, \bibinfo{pages}{142–156}.
\newblock


\bibitem[\protect\citeauthoryear{{Picovoice Inc.}}{{Picovoice Inc.}}{2022}]%
        {porcupine}
\bibfield{author}{\bibinfo{person}{{Picovoice Inc.}}}
  \bibinfo{year}{2022}\natexlab{}.
\newblock \bibinfo{title}{{Porcupine Wake Word Detection \& Keyword Spotting}}.
\newblock
  \bibinfo{howpublished}{\url{https://picovoice.ai/platform/porcupine/}}.
\newblock


\bibitem[\protect\citeauthoryear{{Prebid.org Inc.}}{{Prebid.org Inc.}}{2022}]%
        {prebid}
\bibfield{author}{\bibinfo{person}{{Prebid.org Inc.}}}
  \bibinfo{year}{2022}\natexlab{}.
\newblock \bibinfo{title}{Prebid}.
\newblock \bibinfo{howpublished}{\url{https://prebid.org/}}.
\newblock


\bibitem[\protect\citeauthoryear{{Raspberry Pi}}{{Raspberry Pi}}{2021}]%
        {rpi_access_point}
\bibfield{author}{\bibinfo{person}{{Raspberry Pi}}.}
  \bibinfo{year}{2021}\natexlab{}.
\newblock \bibinfo{title}{{Setting up a Bridged Wireless Access Point}}.
\newblock
  \bibinfo{howpublished}{\url{https://github.com/raspberrypi/documentation/blob/develop/documentation/asciidoc/computers/configuration/access-point-bridged.adoc}}.
\newblock


\bibitem[\protect\citeauthoryear{Ren, Dubois, Choffnes, Mandalari, Kolcun, and
  Haddadi}{Ren et~al\mbox{.}}{2019}]%
        {ren-imc19}
\bibfield{author}{\bibinfo{person}{Jingjing Ren}, \bibinfo{person}{Daniel~J.
  Dubois}, \bibinfo{person}{David Choffnes}, \bibinfo{person}{Anna~Maria
  Mandalari}, \bibinfo{person}{Roman Kolcun}, {and} \bibinfo{person}{Hamed
  Haddadi}.} \bibinfo{year}{2019}\natexlab{}.
\newblock \showarticletitle{Information Exposure From Consumer IoT Devices: A
  Multidimensional, Network-Informed Measurement Approach}. In
  \bibinfo{booktitle}{\emph{Proceedings of the Internet Measurement
  Conference}} \emph{(\bibinfo{series}{IMC '19})}.
  \bibinfo{publisher}{Association for Computing Machinery},
  \bibinfo{address}{Amsterdam}, \bibinfo{pages}{267–279}.
\newblock


\bibitem[\protect\citeauthoryear{{Rhasspy}}{{Rhasspy}}{2022}]%
        {rhasspy}
\bibfield{author}{\bibinfo{person}{{Rhasspy}}.}
  \bibinfo{year}{2022}\natexlab{}.
\newblock \bibinfo{title}{{Rhasspy Voice Assistant}}.
\newblock \bibinfo{howpublished}{\url{https://rhasspy.readthedocs.io/}}.
\newblock


\bibitem[\protect\citeauthoryear{Saidi, Mandalari, Kolcun, Haddadi, Dubois,
  Choffnes, Smaragdakis, and Feldmann}{Saidi et~al\mbox{.}}{2020}]%
        {saidi-imc20}
\bibfield{author}{\bibinfo{person}{Said~Jawad Saidi},
  \bibinfo{person}{Anna~Maria Mandalari}, \bibinfo{person}{Roman Kolcun},
  \bibinfo{person}{Hamed Haddadi}, \bibinfo{person}{Daniel~J. Dubois},
  \bibinfo{person}{David Choffnes}, \bibinfo{person}{Georgios Smaragdakis},
  {and} \bibinfo{person}{Anja Feldmann}.} \bibinfo{year}{2020}\natexlab{}.
\newblock \showarticletitle{{A Haystack Full of Needles: Scalable Detection of
  IoT Devices in the Wild}}. In \bibinfo{booktitle}{\emph{Proceedings of the
  ACM Internet Measurement Conference}} \emph{(\bibinfo{series}{IMC '20})}.
  \bibinfo{publisher}{Association for Computing Machinery},
  \bibinfo{address}{Pittsburgh}, \bibinfo{pages}{87–100}.
\newblock


\bibitem[\protect\citeauthoryear{Shaban}{Shaban}{2018}]%
        {alexa_data_shared_with_stranger}
\bibfield{author}{\bibinfo{person}{Hamza Shaban}.}
  \bibinfo{year}{2018}\natexlab{}.
\newblock \bibinfo{title}{Amazon Alexa user receives 1,700 audio recordings of
  a stranger through `human error'}.
\newblock
  \bibinfo{howpublished}{\url{https://www.washingtonpost.com/technology/2018/12/20/amazon-alexa-user-receives-audio-recordings-stranger-through-human-error/}}.
\newblock


\bibitem[\protect\citeauthoryear{Singh}{Singh}{2019}]%
        {Singh2019VoiceProfiling}
\bibfield{author}{\bibinfo{person}{Rita Singh}.}
  \bibinfo{year}{2019}\natexlab{}.
\newblock \bibinfo{booktitle}{\emph{Profiling humans from their voice}}.
\newblock \bibinfo{publisher}{Springer}.
\newblock


\bibitem[\protect\citeauthoryear{{Software Freedom Conservancy}}{{Software
  Freedom Conservancy}}{2022}]%
        {selenium}
\bibfield{author}{\bibinfo{person}{{Software Freedom Conservancy}}.}
  \bibinfo{year}{2022}\natexlab{}.
\newblock \bibinfo{title}{{Selenium}}.
\newblock \bibinfo{howpublished}{\url{https://www.selenium.dev/}}.
\newblock


\bibitem[\protect\citeauthoryear{{Sonos, Inc.}}{{Sonos, Inc.}}{2022}]%
        {SonosSkill}
\bibfield{author}{\bibinfo{person}{{Sonos, Inc.}}}
  \bibinfo{year}{2022}\natexlab{}.
\newblock \bibinfo{title}{{Sonos}}.
\newblock \bibinfo{howpublished}{\url{https://www.amazon.com/dp/B072ML3N6K}}.
\newblock


\bibitem[\protect\citeauthoryear{{Statista}}{{Statista}}{2022}]%
        {smart_speaker_usage_global}
\bibfield{author}{\bibinfo{person}{{Statista}}.}
  \bibinfo{year}{2022}\natexlab{}.
\newblock \bibinfo{title}{Number of households with smart home products and
  services in use worldwide from 2015 to 2025}.
\newblock
  \bibinfo{howpublished}{\url{https://www.statista.com/statistics/1252975/smart-home-households-worldwide/}}.
\newblock


\bibitem[\protect\citeauthoryear{Statista}{Statista}{2022}]%
        {echo_installed_customer_base}
\bibfield{author}{\bibinfo{person}{Statista}.} \bibinfo{year}{2022}\natexlab{}.
\newblock \bibinfo{title}{Smart speaker devices installed base in the United
  States from 2017 to 2020}.
\newblock
  \bibinfo{howpublished}{\url{https://www.statista.com/statistics/794480/us-amazon-echo-google-home-installed-base/}}.
\newblock


\bibitem[\protect\citeauthoryear{Trimananda, Le, Cui, Ho, Shuba, and
  Markopoulou}{Trimananda et~al\mbox{.}}{2022}]%
        {Trimananda2022Ovrseen}
\bibfield{author}{\bibinfo{person}{Rahmadi Trimananda}, \bibinfo{person}{Hieu
  Le}, \bibinfo{person}{Hao Cui}, \bibinfo{person}{Janice~Tran Ho},
  \bibinfo{person}{Anastasia Shuba}, {and} \bibinfo{person}{Athina
  Markopoulou}.} \bibinfo{year}{2022}\natexlab{}.
\newblock \showarticletitle{{OVRseen}: Auditing Network Traffic and Privacy
  Policies in Oculus {VR}}. In \bibinfo{booktitle}{\emph{31st USENIX Security
  Symposium (USENIX Security 22)}}. \bibinfo{publisher}{USENIX Association},
  \bibinfo{address}{Boston}, \bibinfo{pages}{3789--3806}.
\newblock


\bibitem[\protect\citeauthoryear{{ttm}}{{ttm}}{2022}]%
        {Essential_Oil_Benefits_skill}
\bibfield{author}{\bibinfo{person}{{ttm}}.} \bibinfo{year}{2022}\natexlab{}.
\newblock \bibinfo{title}{{Essential Oil Benefits}}.
\newblock \bibinfo{howpublished}{\url{https://www.amazon.com/dp/B074CNX3G8}}.
\newblock


\bibitem[\protect\citeauthoryear{Varmarken, Le, Shuba, Markopoulou, and
  Shafiq}{Varmarken et~al\mbox{.}}{2020}]%
        {varmarken2020tv}
\bibfield{author}{\bibinfo{person}{Janus Varmarken}, \bibinfo{person}{Hieu Le},
  \bibinfo{person}{Anastasia Shuba}, \bibinfo{person}{Athina Markopoulou},
  {and} \bibinfo{person}{Zubair Shafiq}.} \bibinfo{year}{2020}\natexlab{}.
\newblock \showarticletitle{{The TV is Smart and Full of Trackers: Measuring
  Smart TV Advertising and Tracking}}. In \bibinfo{booktitle}{\emph{Proceedings
  on Privacy Enhancing Technologies}}, Vol.~\bibinfo{volume}{2020 (2)}.
  \bibinfo{publisher}{Sciendo}, \bibinfo{address}{Virtual},
  \bibinfo{pages}{129--154}.
\newblock


\bibitem[\protect\citeauthoryear{{VCA, INC.}}{{VCA, INC.}}{2022}]%
        {VCAAnimalHospitalSkill}
\bibfield{author}{\bibinfo{person}{{VCA, INC.}}}
  \bibinfo{year}{2022}\natexlab{}.
\newblock \bibinfo{title}{{VCA Animal Hospital}}.
\newblock \bibinfo{howpublished}{\url{https://amazon.com/dp/B07KYS1Y1X}}.
\newblock


\bibitem[\protect\citeauthoryear{{WaLLy3K}}{{WaLLy3K}}{2022}]%
        {PiHole}
\bibfield{author}{\bibinfo{person}{{WaLLy3K}}.}
  \bibinfo{year}{2022}\natexlab{}.
\newblock \bibinfo{title}{{Pi-hole Blocklist}}.
\newblock \bibinfo{howpublished}{\url{https://firebog.net/}}.
\newblock


\bibitem[\protect\citeauthoryear{Willens}{Willens}{2021}]%
        {amazon_identity}
\bibfield{author}{\bibinfo{person}{Max Willens}.}
  \bibinfo{year}{2021}\natexlab{}.
\newblock \bibinfo{title}{Amid post-cookie confusion, Amazon plans to launch an
  identifier of its own}.
\newblock
  \bibinfo{howpublished}{\url{https://digiday.com/marketing/amid-post-cookie-confusion-amazon-explores-launching-an-identifier-of-its-own/amp/}}.
\newblock


\bibitem[\protect\citeauthoryear{{Xeline Development}}{{Xeline
  Development}}{2022}]%
        {makeup-of-the-day-skill}
\bibfield{author}{\bibinfo{person}{{Xeline Development}}.}
  \bibinfo{year}{2022}\natexlab{}.
\newblock \bibinfo{title}{{Makeup of the Day}}.
\newblock \bibinfo{howpublished}{\url{https://amazon.com/dp/B072N6BNB1}}.
\newblock


\bibitem[\protect\citeauthoryear{Young, Liao, Cheng, Hu, and Deng}{Young
  et~al\mbox{.}}{2022}]%
        {Young22SkillDetectiveUSENIX}
\bibfield{author}{\bibinfo{person}{Jeffrey Young}, \bibinfo{person}{Song Liao},
  \bibinfo{person}{Long Cheng}, \bibinfo{person}{Hongxin Hu}, {and}
  \bibinfo{person}{Huixing Deng}.} \bibinfo{year}{2022}\natexlab{}.
\newblock \showarticletitle{{SkillDetective}: Automated {Policy-Violation}
  Detection of Voice Assistant Applications in the Wild}. In
  \bibinfo{booktitle}{\emph{31st USENIX Security Symposium (USENIX Security
  22)}}. \bibinfo{publisher}{USENIX Association}, \bibinfo{address}{Boston},
  \bibinfo{pages}{1113--1130}.
\newblock


\bibitem[\protect\citeauthoryear{{YouVersion}}{{YouVersion}}{2022}]%
        {YouVersionBibleSkill}
\bibfield{author}{\bibinfo{person}{{YouVersion}}.}
  \bibinfo{year}{2022}\natexlab{}.
\newblock \bibinfo{title}{{YouVersion Bible}}.
\newblock \bibinfo{howpublished}{\url{https://www.amazon.com/dp/B017RXFNKY}}.
\newblock


\bibitem[\protect\citeauthoryear{Zeng, Kohno, and Roesner}{Zeng
  et~al\mbox{.}}{2021}]%
        {zeng2021makes}
\bibfield{author}{\bibinfo{person}{Eric Zeng}, \bibinfo{person}{Tadayoshi
  Kohno}, {and} \bibinfo{person}{Franziska Roesner}.}
  \bibinfo{year}{2021}\natexlab{}.
\newblock \showarticletitle{What Makes a “Bad” Ad? User Perceptions of
  Problematic Online Advertising}. In \bibinfo{booktitle}{\emph{Proceedings of
  the 2021 CHI Conference on Human Factors in Computing Systems}} (Yokohama,
  Japan) \emph{(\bibinfo{series}{CHI '21})}. \bibinfo{publisher}{Association
  for Computing Machinery}, \bibinfo{address}{New York, NY, USA}, Article
  \bibinfo{articleno}{361}, \bibinfo{numpages}{24}~pages.
\newblock
\showISBNx{9781450380966}
\urldef\tempurl%
\url{https://doi.org/10.1145/3411764.3445459}
\showDOI{\tempurl}


\bibitem[\protect\citeauthoryear{Zhang, Psounis, Haroon, and Shafiq}{Zhang
  et~al\mbox{.}}{2022}]%
        {zhang2021harpo}
\bibfield{author}{\bibinfo{person}{Jiang Zhang}, \bibinfo{person}{Konstantinos
  Psounis}, \bibinfo{person}{Muhammad Haroon}, {and} \bibinfo{person}{Zubair
  Shafiq}.} \bibinfo{year}{2022}\natexlab{}.
\newblock \showarticletitle{HARPO: Learning to Subvert Online Behavioral
  Advertising}. In \bibinfo{booktitle}{\emph{29th Annual Network and
  Distributed System Security Symposium}}. \bibinfo{publisher}{Internet
  Society}, \bibinfo{address}{San Diego}.
\newblock


\end{thebibliography}

\section*{Appendix}
\label{sec:appendix}
\renewcommand{\thesubsection}{\Alph{subsection}}

\subsection{High absolute bid values with only skill installation}
\label{appendix:high-bid-values}
This Appendix expands on the discussion of ad bid values after skill installation discussed in Section~\ref{sub-sec:interaction-bidding}.
We note that the absolute bid values for vanilla and interest personas with only skill installation are higher than that of vanilla and interest personas with user interaction. 
While it is impossible to know the exact reason for high bid values for personas with only skill installation, a possible explanation could be our data collection during the holiday season, i.e., around Christmas 2021.\footnote{We collected advertisements (including bids) both before and after interacting with the skills from 12/08/21 to 12/22/21 and 12/28/21 to 02/04/22, respectively. We installed the skills on 12/07/21 and we interacted with the skills on 12/27/21.}
To rule out the impact of holiday season, we compare the bids values with only skill installation and with skill interaction that were collected close to each other. 
Specifically, we compare the bids from last three iteration of without interaction with bids from first three iterations of with interaction, that were crawled within a close time span.\footnote{From 12/20/21 to 12/22/21 for skill installation and from 12/28/21 to 01/12/22 for skill interaction.}
Table \ref{table:bids-mean-last-no-interaction-first-interaction} presents mean bid values without and with user interaction. 
It can be seen that the interest personas with interaction receive higher bids than vanilla persona.
Whereas no discernible differences exist for without interaction configurations.

\begin{table}[t]
    \centering
    \small
    \begin{tabular}{>{\bfseries}l?{0.3mm}cc}
    \toprule
    \textbf{Persona}            & \textbf{No Interaction} & \textbf{Interaction} \\
    \midrule
    Connected Car               & 0.364 & 0.311 \\
    Dating                      & 0.519 & 0.297 \\
    Fashion \& Style            & 0.572 & 0.404 \\
    Pets \& Animals             & 0.492 & 0.373 \\
    Religion \& Spirituality    & 0.477 & 0.231 \\
    Smart Home                  & 0.452 & 0.349 \\
    Wine \& Beverages           & 0.418 & 0.522 \\
    Health \& Fitness           & 0.564 & 0.826 \\
    Navigation \& Trip Planners & 0.533 & 0.268 \\
    \midrule
    Vanilla                     & 0.539 & 0.232 \\
    \bottomrule
    \end{tabular}
    \caption{Mean bid values without and with interaction across interest and vanilla personas that were collected close to each other.}
    \label{table:bids-mean-last-no-interaction-first-interaction}
\end{table}

Although the timing affects the bid values, we believe that it does not impact our findings.
Specifically, our objective is to measure the effect of treatment, i.e., skill installation or interaction, on interest (treatment) personas as compared to the vanilla (control) persona. 
The relative comparison of bid values between vanilla (control) and interest (treatment) personas suffices to measure the effect of treatment.
It means that, if we see statistically significant differences in bid values between vanilla (control) and interest (treatment) personas, we can confidently attribute the differences to the applied treatment, i.e., skill installation or interaction.

\end{document}